\documentclass[9pt,twocolumn,twoside,lineno]{pnas-new}
\usepackage{textcomp}
\usepackage{physics}

\newcommand{\ad}{\hat{a}^{\dag}}

\newcommand{\ir}{\mathrm{i}}

\newcommand{\gc}{g_{\mathrm{c}}}

\newcommand{\muc}{\mu_{\mathrm{c}}}
\newcommand{\muf}{\mu_{\mathrm{f}}}
\newcommand{\omi}{\omega_{0,\lambda}}
\newcommand{\omf}{\omega_{\mathrm{f},\lambda}}

\templatetype{pnasresearcharticle} 
\setboolean{displaywatermark}{false}

\title{Metastability and discrete spectrum of long-range systems}

\author[a]{Nicol\`o Defenu}

\affil[a]{Institut f\"ur Theoretische Physik, ETH Z\"urich, Wolfgang-Pauli-Str.\,27, 8093 Z\"urich, Switzerland}
\leadauthor{Nicol\`o Defenu} 

\significancestatement{The linear dynamics of closed quantum systems produces well known difficulties in the definition of quantum chaos. This leads to several issues in the theoretical justification of the equilibration and thermalisation dynamics observed in closed experimental systems. In the case of large harmonic baths these issues are partially resolved due to the continuous nature of the spectrum, which produces divergent Poincar\'e recurrence times. As it is shown in the following, such scenario does not apply to long-range interacting models, whose spectrum remains discrete up to the thermodynamic limit, in contradiction with the textbook description of infinitely extended systems. }

\authorcontributions{Nicol\`o Defenu is the sole author of the paper.}
\authordeclaration{The author declares no competing interests.}
\correspondingauthor{\textsuperscript{1} E-mail: ndefenu@phys.ethz.ch}


\begin{abstract}
Long lived quasi-stationary states (QSSs) are a signature characteristic of long-range interacting systems both in the classical and in the quantum realms. Often, they emerge after a sudden quench of the Hamiltonian internal parameters and present a macroscopic life-time, which increases with the system size. Despite their ubiquity, the fundamental mechanism at their root remains unknown. Here, we show that the spectrum of systems with power-law decaying couplings remains discrete up to the thermodynamic limit. As a consequence, several traditional results on the chaotic nature of the spectrum in many-body quantum systems are not satisfied in presence of long-range interactions. In particular, the existence of QSSs may be traced back to the finiteness of Poincar\'e recurrence times. This picture justifies and extends known results on the anomalous magnetization dynamics in the quantum Ising model with power-law decaying couplings. The comparison between the discrete spectrum of long-range systems and more conventional examples of pure point spectra in the disordered case is also discussed. \end{abstract}

\dates{This manuscript was compiled on \today}
\doi{\url{https://doi.org/10.1073/pnas.2101785118}}

\begin{document}

\maketitle
\thispagestyle{firststyle}
\ifthenelse{\boolean{shortarticle}}{\ifthenelse{\boolean{singlecolumn}}{\abscontentformatted}{\abscontent}}{}

\dropcap{E}quilibration is at the roots of thermodynamics and has been verified under general conditions in a wide range of physical systems. The current scientific literature has focused on several aspects of this problem, starting from quantum quenches and relaxation\,\cite{cramer2008exact,polkovnikov2011colloquium}, and arriving to thermalization of integrable and quasi-integrable systems\,\cite{kinoshita2006quantum,rigol2009breakdown,relanon2010thermalization}, typicality as a foundation of quantum statistical mechanics \cite{reimann2008foundation,linden2009quantum,goldstein2010approach} and many others.

Despite the ubiquity of equilibration, or possibly due to it, the known examples of diverging equilibration times and recurrent behaviour have attracted wide attention in modern physics. Dynamical protocols, where the system initially relaxes into a long-lived states and, then, undergoes actual equilibration on a
longer timescale, are commonly referred to as metastable. The observation of various dynamical regimes, separated by distinct timescales is the most common evidence of metastability and is found in several classical systems and, especially, glasses\,\cite{binder2005glassy,angelini2011glass,biroli2013perspective}. In  closed quantum systems ordinary examples of metastability appear in presence of localised states\,\cite{affleck1981quantum,polkovnikov2011colloquium,eisert2015quantum}, whose energy eigenvalues are separated from the rest of the spectrum.  An analogous spectral structure justifies the observation of metastability also in open quantum systems\,\cite{macieszczak2016towards}.  

Diverging equilibration times in the thermodynamic limit are also a notorious characteristic of long-range interacting systems. A physical system is said to be long-range when the two-body interaction potential decays as a power law of the distance $r$ between its microscopic components: $V(r)\sim r^{-\alpha}$ in the large distance ($r\to\infty$) limit. If one focuses on the thermodynamic behaviour, two main regimes appear as a function of $\alpha$. For $\alpha>d$, where $d$ is the spatial dimension of the system, textbook thermodynamics is well defined and long-range interactions only alter the universal scaling close to critical points\,\cite{defenu2020criticality}. We refer to this regime ($\alpha>d$) as weak long-range interactions.

Conversely, for $\alpha<d$ the thermodynamic quantities become non-additive, leading to apparently paradoxical predictions such as ensemble in-equivalence or negative specific heat and susceptibilities\,\cite{campa2009statistical}. In the out-of-equilibrium realm, the most striking property of strong long-range systems is the appearance of quasi-stationary states (QSS), i.e. metastable configurations whose lifetime scales super-linearly with the system size. QSSs have been mainly studied in classical systems, such as the Hamiltonian Mean-Field model\,\cite{antoni1995clustering}, where an ensemble of plane rotators are subject to a fully connected flat interaction ($\alpha=0$). There, QSSs are often described in terms of the magnetisation dynamics, which, after a sudden quench from an appropriate set of initial conditions, stabilises to a different value with respect to its equilibrium expectation. Then, actual equilibration only occurs after a macroscopic time-scale $\tau\propto N^{\beta}$ with $\beta>0$\,\cite{campa2009statistical}.
Apart from this peculiar case, QSSs are characteristic of long-range interactions\,\cite{gabrielli2010quasistationary}, ranging from gravitational\,\cite{joyce2010relaxation} to electromagnetic systems\,\cite{gupta2017world}. 

The advent of cold atom experiments has largely broadened the interest in long-range physics, due to the possibility of realising non-local interactions via several different means, such as dipolar systems\,\cite{dell2000bose,micheli2006toolbox,baranov2012condensed}, cold atoms excited into Rydberg states\,\cite{saffman2010quantum} and trapped ions\,\cite{monroe2020programmable}. In the context of meta-stable dynamics and QSSs a crucial role is played by cold atoms confined in optical resonators, where the photons are stored within the cavity for a sufficiently long time and mediate interactions whose range extends over the entire cavity volume\,\cite{munstermann2000observation}. At the semi-classical level, a strict relation between the dynamics of cold atoms into cavity systems and the one of the Hamiltonian Mean-Field model has been demonstrated\,\cite{schutz2016dissipation}, designating these devices as optimal candidates for the observation of slow or absent equilibration.

Given this broad physical interest, as well as the universal presence of QSSs in long-range interacting systems, it is surprising that the general mechanism at the root of their existance has still to be identified. Indeed,
while most results concerning QSSs in classical systems derive from numerical simulations\,\cite{campa2009statistical}, first evidences of their appearance in the quantum realm have been rooted on an analytic approach, which was, however, restricted to a  $1/2$-spin Hamiltonian with specific boundaries of the quench protocol\,\cite{kastner2011diverging}.

In the present manuscript, we are going to prove that the absence of equilibration of long-range quantum systems is directly connected to the persistence of finite Poincar\'e recurrence times also in the thermodynamic limit. Hence, the physics of macroscopic long-range systems cannot be described by the ``traditional" thermodynamic limit procedure. This is in agreement with well-known observations of properties, which are common to thermodynamically large long-range systems and finite local ones, such as the impossibility to fully disregard boundary over bulk phenomena\,\cite{barre2007ensemble,latella2015thermodynamics}, the existence of concave entropy regions\,\cite{ispolatov2001first} or the presence of a macroscopic energy gap between the ground state and the first excited state\,\cite{gupta2012one,gupta2012overdamped}. 

All the above features are consequences of the spectral properties of  long-range many-body systems, whose spectrum does not become continuous in the thermodynamic limit. Indeed, the eigenvalues of a long-range coupling matrix can be shown to remain discrete even in the infinite components limit, forming a pure point spectrum\,\cite{last1996quantum} similar to the one observed in celebrated examples of disordered systems\,\cite{thouless1972anderson}. However, at variance with most disordered systems, the spectrum of strong long-range interactions possesses no continuous subspace in the thermodynamic limit, in analogy with the case of the Anderson model at infinite disorder strength\,\cite{froehlich1983absence,simon1985some,scardicchio2017perturbation}. 

The paper is organised as follows: In Sec.\,\ref{hth_chaos} we are going to outline the general picture for equilibration in closed integrable quantum systems. In Sec.\,\ref{spec_lr_systems} a proof of the spectral discreteness of long-range couplings in the thermodynamic limit is presented in the case of the tight binding chain. Then, in Sec.\,\ref{rec_time}, the  connection between this result and the vanishing of the Poincar\'e recurrence times for critical quenches in generic quantum system is explored. In Sec.\,\ref{kit_ch}, the above picture is employed to justify the observation of diverging equilibration times in a long-range Ising model, quenched across its quantum critical point\,\cite{kastner2011diverging}. In Sec.\,\ref{sph_model} lack of equilibration is shown in an ensemble of long-range coupled spin-waves for a generic (non critical) quench. Finally, in Sec.\,\ref{disc} the conclusive remarks are reported. 

\section{H-theorem and kinematical chaos}
\label{hth_chaos}
The divergence of the recurrence times for thermodynamically large classical systems was already noticed by Boltzmann\,\cite{boltzmann1896entgegnung} in answering Zermelo's criticism\,\cite{zermelo1896ueber} to the H-theorem (see Ref.\,\cite{steckline1983zermelo} for an historical account). Quite interestingly, a similar dispute has successively arisen for quantum systems, where the issue of recurrence is more severe with respect to the classical case\,\cite{hogg1982recurrence}. There, the coarse grained entropy was  shown to be a quasi-periodic function and the validity of H-theorem was questioned\,\cite{percival1961almost}. Eventually, these observations were proved to be inconsequential for macroscopic quantum systems, where the wave-function recurrence times become exponentially large\,\cite{peres1982recurrence} in the thermodynamic limit. Hence, the validity of the H-theorem is recovered\,\cite{percival1961almost}.

The issue of recurrence times in quantum systems is profoundly tied with the mathematical theory of quasi-periodic functions\,\cite{bohr1952collected,besicovitch1954almost}. This connection can be concretely explored by considering a system with time-independent Hamiltonian $\hat{H}$ initially prepared at $t=0$ in a pure state $|\psi\rangle$, which does not belong to the Hamiltonian spectrum. As long as the system is bounded, i.e. has a finite volume, the spectrum is discrete and the Hamiltonian can be decomposed in terms of orthogonal projectors $\hat{\Pi}_{n}$
\begin{align}
\hat{H}=\sum_{n}E_{n}\hat{\Pi}_{n}
\end{align}
which define the states
\begin{align}
|n\rangle=\frac{\hat{\Pi}_{n}|\psi\rangle}{||\hat{\Pi}_{n}|\psi\rangle||^{2}}.
\end{align}
Accordingly, any dynamical observable can be represented as sum of time oscillating functions in the $|n\rangle$ states basis.

Equilibration is conveniently quantified by the fidelity 
\begin{align}
\label{fid}
f(t)=||\langle \psi |e^{-i\hat{H}t}|\psi\rangle||^{2}=|\chi(t)|^{2}
\end{align}
which represents the overlap between the initial state $|\psi\rangle$ and its time evolution $|\psi (t)\rangle=e^{-i\hat{H}t}|\psi\rangle$. The fidelity is obtained as the square of the characteristic function
\begin{align}
\label{char_fun}
\chi(t)=\sum_{n}p_{n}e^{-itE_{n}}\quad \mathrm{with}\quad p_{n}=\langle\psi|\hat{\Pi}_{n}|\psi \rangle.
\end{align}
It can be proven that the sum in \eqref{char_fun} yields an almost periodic function and, thus, the time evolved state will periodically return arbitrarily close to the initial state\,\cite{besicovitch1954almost}. As the system size grows, approaching the thermodynamic limit, at least some portions of the spectrum are expected to become absolutely continuous and
\begin{align}
\label{rl_lemma}
\lim_{t\to \infty}\chi(t)=0
\end{align} 
due to the Riemann--Lebesgue lemma\,\cite{venuti2015recurrence}.

 From the more general perspective of spectral theory, any physical Hilbert space $\mathcal{H}$ may be decomposed into an absolutely continuous subspace $\mathcal{H}_{\rm ac}$, a singular continuous subspace $\mathcal{H}_{\rm sc}$ and a pure point subspace $\mathcal{H}_{\rm pp}$ ($\mathcal{H}=\mathcal{H}_{\rm ac}\oplus\mathcal{H}_{\rm sc}\oplus\mathcal{H}_{\rm pp}$). For any initial state $|\psi\rangle$ with vanishing projection on the pure point subspace of the spectrum, it can be demonstrated that 
 \begin{align}
\label{rage_th}
\lim_{T\to\infty}\langle f(t) \rangle_{T}=0,\quad\mathrm{where}\quad \langle [...] \rangle_{T}=\frac{1}{T}\int_{0}^{T}[...]dt
\end{align}
is the Cesaro's time average\,\cite{last1996quantum}\footnote{ A similar result holds for the expectation value of any compact operator\,\cite{last1996quantum}.}. These results indicate that in closed quantum systems with continuous spectra equilibration occurs due to the decoherence of dynamical fluctuations, either in the absolute sense, see \eqref{rl_lemma}, or on average, see \eqref{rage_th}.

Furthermore, most Hamiltonians, which possess a finite pure point subspace in the spectrum, occur in the field of disordered systems, where localised states appear whose energy lies outside the continuous bulk band\,\cite{froehlich1983absence,simon1985some,scardicchio2017perturbation}. There, metastability appears for those initial states whose projection in the pure point region of the spectrum  hinders the applicability of \eqref{rage_th}. In the following, we are going to demonstrate that the spectrum of translational invariant long-range systems is completely constituted by a pure point subspace (with a single accumulation point at high energy). Therefore, metastability in long-range systems occurs for any initial state as the conditions for \eqref{rage_th} are never verified. 

In summary, most quantum many-body systems are expected to equilibrate following roughly the same chaotic behaviour of classical systems and the result in \eqref{rl_lemma} has been referred to as \emph{kinematical chaos}\,\cite{lasinio1996chaotic,reimann2008foundation}. The aforementioned scenario has been explicitly verified in several solvable quantum models\,\cite{emch1966non,goldstein1974space,radin1970approach,lenci1996ergodic} and is one of the fundamental assumptions of the eigenstate thermalisation hypothesis (ETH) in non-integrable quantum many-body systems\,\cite{srednicki1994chaos, rigol2008thermalization}. However, this scenario does not hold for non-additive long-range systems.

\section{Spectrum of long-range systems}
 \label{spec_lr_systems}
In the following, the lack of equilibration evidenced in long-range quantum systems\,\cite{kastner2011diverging, bachelard2013universal,eisert2013breakdown,metivier2014spreading,rajabpour2015quantum,mori2017classical} is shown to derive from the breakdown of the \emph{kinematical chaos} hypothesis. We consider a generic Hamiltonian with long range translational invariant couplings in one dimension
\begin{align}
\label{h1}
\hat{H}=-\sum_{i=1}^{N}\sum_{r=1}^{N/2-1}t_{r}(\hat{a}^{\dagger}_{i}\hat{a}_{i+r}+h.c.)+\mu\sum_{i=1}^{N}\hat{a}_{i}^{\dagger}\hat{a}_{i}+\hat{H}_{\rm int},\end{align}
where the $\hat{a}_{i}^{\dagger}(\hat{a}_{i})$ symbols represent the creation(annihilation) operators of quantum particles on the $i-th$ site of the chain and $N$ is the total number of sites. The bosonic or fermionic nature of the particles nor the specific shape of the interaction Hamiltonian $\hat{H}_{\rm int}$ are crucial to our arguments. 

The long-range hopping amplitudes take the form,
\begin{align}
\label{hop_am}
t_{r}=\frac{1}{N_{\alpha}}\frac{1}{r^{\alpha}},
\end{align}
where the factor $N_{\alpha}=\sum_{r=1}^{N/2}r^{-\alpha}$ has to be introduced in order to yield an extensive internal energy\,\cite{kac1963van}. In the large size limit the scaling term reads
\begin{align}
\label{k_norm}
N_{\alpha}^{-1}\approx\begin{cases}
(1-\alpha)2^{(1-\alpha)}N^{\alpha-1}&\quad\mathrm{if}\,\,\alpha<1\\
1/\log(N)&\quad\mathrm{if}\,\,\alpha=1\\
1/\zeta(\alpha)&\quad\mathrm{if}\,\,\alpha>1.
\end{cases}
\end{align}
In general, the spectrum of any interacting Hamiltonian in \eqref{h1} can be obtained by means of perturbation theory\,\cite{messiah1961quantum}. Then, the first step is to diagonalise the non-interacting Hamiltonian
\begin{align}
\label{h0}
\hat{H}_{0}=-\sum_{i=1}^{N}\sum_{r=1}^{N/2-1}t_{r}(\hat{a}^{\dagger}_{i}\hat{a}_{i+r}+h.c.)+\mu\sum_{i=1}^{N}\hat{a}_{i}^{\dagger}\hat{a}_{i}.\end{align}
Assuming periodic boundary conditions at the edge of the chains $a_{i+L}=a_{i}\,\,(a^{\dagger}_{i+L}=a_{i}^{\dagger})$, the spectrum of the non-interacting Hamiltonian $H_{0}$ is obtained as $\varepsilon(k)=\mu-\tilde{t}_{k}$, where
\begin{align}
\label{ft}
\tilde{t}_{k}=\sum_{r=1}^{N/2-1}\cos(kr)t_{r}=\frac{1}{N_{\alpha}}\sum_{r=1}^{N/2-1}\frac{\cos(kr)}{r^{\alpha}}
\end{align}
is the Fourier coefficient of the hopping amplitudes $t_{r}$ in \,\eqref{hop_am}.  The periodic boundary conditions impose the usual restriction on the particle momentum $k\equiv k_{n}=2\pi n/N$ with $n\in \mathbb{Z}$ and $-N/2\leq n<N/2$ (the lattice spacing has been set to $1$). As long as $\alpha>
1$ the calculation proceeds as in the nearest-neighbour case and the thermodynamic limit of \eqref{ft} can be taken safely, substituting the discrete momentum values $k_{n}$ with the continuous value $k\in[-\pi,\pi)$. Accordingly, the spectrum of the Hamiltonian for $\alpha>1$ becomes continuous and the \emph{kinematical chaos} hypothesis applies.

Conversely, for $\alpha<1$ the Kac normalization factor $N_{\alpha}$ in \eqref{k_norm} diverges at large $N$ and the thermodynamic limit  of \eqref{ft} has to be carefully considered. Therefore, it is convenient to write explicitly \eqref{ft} at large $N$
\begin{align}
\label{asy_ft}
\lim_{N\to\infty}\frac{1}{N_{\alpha}}\sum_{r=1}^{N/2-1}\frac{\cos(kr)}{r^{\alpha}}\approx\frac{c_{\alpha}}{N}\sum_{r=1}^{N/2}\frac{\cos\left(2\pi n\frac{r}{N}\right)}{(r/N)^{\alpha}}
\end{align}
where the asymptotic form of the Kac normalisation in \eqref{k_norm} has been employed and, accordingly, the size independent constant reads $c_{\alpha}=(1-\alpha)2^{1-\alpha}$. Thanks to the $1/N$ scaling of the discrete momenta on the lattice, the summation in \eqref{asy_ft} only depends on the variable $r/N$ and in the $N\to\infty$ limit the Riemann summation formula can be applied\,\cite{hughes2008calculus}
\begin{align}
\label{core_res}
\tilde{t}_{n}\equiv\lim_{N\to\infty}\tilde{t}_{k}=c_{\alpha}\int_{0}^{\frac{1}{2}}\frac{\cos\left(2\pi n\,s\right)}{s^{\alpha}}ds.
\end{align} 
Despite its simplicity, the result in \eqref{core_res} has profound physical implications: it proves that the spectrum of a quantum system with long-range harmonic couplings remains discrete also at $N\to\infty$. Indeed, at $\alpha<1$ the gap between neighbouring eigenvalues $\omega_{n+1}-\omega_{n}$ labeled by the consecutive momenta $k_{n},\,k_{n+1}$ in \eqref{ft} does not vanish in the thermodynamic limit as it would for $\alpha>1$. As a consequence, the energy eigenvalues only depend on the integer index $n\in \mathbb{Z}$ rather than on the continuous momentum $k$
\begin{align}
\label{disc_spec}
\omega_{n}=\mu-\tilde{t}_{n}.
\end{align}
Notably, for $\alpha=0$ all the $\tilde{t}_{n}$ coefficient vanish and the discrete spectrum in \eqref{disc_spec} becomes fully degenerate, which is often referred to as an essential spectrum\,\cite{last1996quantum}. 

From the perspective of disordered systems the result in \eqref{disc_spec} is rather remarkable. Indeed, the spectrum of long-range couplings in the thermodynamic limit is a completely pure point spectrum with no continuous subspace.  The same situation is found in an Anderson model with infinite disorder strength, where localisation occurs at all energy scales\,\cite{froehlich1983absence,simon1985some,scardicchio2017perturbation}. However, most examples of disordered systems feature both continuous and pure point subspaces in the spectrum. Accordingly, metastable configurations appears in disordered systems only for states with finite projection on the pure point subspace. This is not the case of long-range systems where lack of equilibration is a ubiquitous feature of all possible initial states. Furthermore, it is worth noting that the energies in \eqref{disc_spec} are not dense, differently from other examples in disordered systems\,\cite{froehlich1983absence,simon1985some,scardicchio2017perturbation}. Rather, any energy eigenvalue in \eqref{disc_spec} is isolated and the only accumulation point occurs at the maximum energy $\max_{n}\omega_{n}=\mu$, since $\lim_{n\to\infty}\tilde{t}_{n}\to 0$ as follows from  the Riemann--Lebesgue lemma\,\cite{last1996quantum}.

It is evident that the core result in \eqref{core_res} would not be altered by the nature of the particles (bosons or fermions) nor by most interaction terms $\hat{H}_{\rm int}$. A simple argument to substantiate the above claim may be obtained by an inspection of the perturbative corrections for the eigenvalues of the $\hat{H}$ Hamiltonian caused by the interaction term $\hat{H}_{\rm int}$
\begin{align}
\label{pt_res}
\delta E_{n}=\langle \psi_{n} | \hat{H}_{\rm int} | \psi_{n} \rangle+\sum_{n\neq n'}\frac{|\langle \psi_{n} | \hat{H}_{\rm int} | \psi_{n'} \rangle|^{2}}{E_{n}-E_{n'}} +\cdots,
\end{align}
where $| \psi_{n} \rangle$ are the symmetric(antisymmetric) external product of the single particle eigenstates $| k_{n} \rangle$ of the periodic chain and $E_{n}$ their energy. 

As long as the system is finite, the spectrum can be safely assumed to be discrete and non-degenerate, so that the perturbative result in \eqref{pt_res} will yield a good approximation to the $\hat{H}$ spectrum for weak enough perturbations. Conventionally, one expects similar perturbative arguments to breakdown in the thermodynamic limit, due to the possible divergent contributions arising from fluctuations close to critical points. However, this is not the case of strong long-range systems, where the long-range tails of the couplings are known to suppress strong fluctuations\,\cite{mukamel2009notes}. Indeed the discreteness of the non-interacting many-body spectrum, see \eqref{core_res}, persists in the thermodynamic limit and the perturbative contributions as the ones on the r.h.s. of \eqref{pt_res} shall not develop any singularity. 

Therefore, the discreteness of the spectrum, derived in \eqref{core_res}, may be presumed to persist in most interacting Hamiltonians, since one of its main consequences is to suppress strong interaction contributions in perturbation theory. Hence, the physics of thermodynamically large long-range systems would be closer to the one of a finite bounded Hamiltonians, rather than to the one of traditional many-body ensembles. The generalisation of the result in \eqref{core_res} to the $d>1$ case follows the same procedure indicated in $1$-dimension, as it is shown in the SI Appendix. 

\section{Vanishing recurrence time in the \texorpdfstring{$N\to\infty$}\, limit}
\label{rec_time}
 
Previous section clarified that the spectrum of quantum long-range systems remains discrete in the thermodynamic limit. It follows that the recurrence times for these systems remain finite\,\cite{percival1961almost} and that the hypothesis of \emph{kinematical chaos} does not apply to power-law decaying couplings with $\alpha<d$. An analogous situation occurs in the low-energy limit of the Anderson model where the spectrum is also pure point\,\cite{froehlich1983absence,simon1985some,scardicchio2017perturbation}. 

Yet, this is not enough to justify the size scaling of the QSSs lifetimes, which actually diverge in the thermodynamic limit\,\cite{dauxois2002dynamics,kastner2011diverging}. Rather, this effect is the result of the accumulation of the energy levels in \eqref{core_res} towards the high-energy ($n\to\infty$) limit. In order to substantiate this argument, let us revisit the calculation of the recurrence time for a discrete spectrum\,\cite{slater1939rates,peres1982recurrence,bhattacharyya1986estimates} in the light of the result in \eqref{core_res}. Then, following Ref.\,\cite{bhattacharyya1986estimates}, one writes  the fidelity in \eqref{fid} as $f(t)=1-Q(t)$, so that a recurrence will be achieved each time $Q(t)\simeq 0$. It is convenient to assume that only a finite number of states $M$ contributes to \eqref{char_fun} and all have the same population  $p_{n}=1/M$, yielding the simplified result
\begin{align}
\label{sub_fid}
Q(t)=\frac{4}{M^{2}}\sum_{m>n=1}^{M}\sin^{2}\left(\frac{\omega_{nm}\,t}{2}\right),
\end{align}
where $\omega_{nm}=\omega_{m}-\omega_{n}$ is the difference between the two energy eigenstates.
Then, one can repeat the arguments of Ref.\,\cite{peres1982recurrence} and relate the smallest recurrence time $\tau$ with the probability that a cylinder in $M-1$-dimensional space contains at least one point of a regular $M-1$-dimensional lattice. The cylinder radius is $R\approx \sqrt{(M-1)\varepsilon/8}$, with $\varepsilon$ any small parameter such that $Q(t)<\varepsilon$, while the  length is proportional to the recurrence time itself $L=\sqrt{M-1}\omega \tau$, multiplied by the average square frequency
\begin{align}
\label{av_lev_spac}
\omega=\sqrt{\frac{1}{M-1}\sum_{m=2}^{M}\omega_{1m}^{2}}.
\end{align}
In order for the cylinder to contain at least one  point of the regular $M-1$-dimensional lattice, its volume has to be approximately equal to one, i.e. $ \sqrt{M-1}\omega \tau\sigma(R)\approx 1$, where $\sigma(R)$ is the volume of a $M-2$ dimensional sphere of radius $R$, leading to
\begin{align}
\label{rec_time_est}
\tau=\frac{1}{\sqrt{M-1}\omega\sigma(R)}.
\end{align}
Apart from the details for the derivation of \eqref{rec_time_est}, which can be found in Refs.\,\cite{slater1939rates,peres1982recurrence}, its interpretation is rather evident: the scale for the recurrence time is set by the average level spacing $\omega$, see \eqref{av_lev_spac}, but the net result is inversely proportional to the volume of the $(M-2)$-dimensional sphere of radius $R$. Such volume vanishes as the accuracy requested for the recurrence time is increased, i.e. $\varepsilon\to 0$, but also in the limit of infinitely many energy levels involved ($M\to \infty$). Indeed, within the conventional assumption of rational independence between the energy eigenvalues\,\cite{percival1961almost}, each novel energy eigenstate introduces an independent direction to the space of $\omega_{1m}$ and lowers the probability to find a common recurrence time for the entire spectrum.

In general, the number of states available for a quantum system grows with its size and accumulates at low energies. Then, the result in \eqref{rec_time_est} diverges, as $\sigma(R)$ vanishes in the $M\to\infty$ limit. However, based on the result in \eqref{core_res}, the energy eigenvalues of long-range systems do not densely fill a continuous band. Rather as $N$ grows the eigenstates accumulate and become dense at high-energy, due to the growing index of Fourier modes $-N/2<n<N/2$. Indeed, the energy difference between these modes becomes increasingly negligible at large $n$, since the integral in \eqref{core_res} approaches zero in the $n\to\infty$ limit. Therefore, the conventional assumption of rational independence for the energy levels in \eqref{sub_fid} does not apply at large $M$ as most of the states at high-energy are quasi-degenerate. Accordingly, one may expect the volume of the $M$-sphere in \eqref{rec_time_est} to stop growing for $M$ above a certain (unknown) threshold $M^{*}$. Within this assumption, the formula in \eqref{rec_time_est} at large $M$ becomes
\begin{align}
\label{rec_time_est_large_N}
\lim_{M\to\infty}\tau\sim\frac{1}{\sqrt{M}\omega\sigma_{\rm max}}
\end{align}
which vanishes in the large-$M$ limit.

 In summary, one may generally expect that the number of levels $M$ involved in the computation of the recurrence time $\tau$, see \eqref{rec_time_est}, grows with the size of the system $N$. However, differently from the standard case\,\cite{slater1939rates,peres1982recurrence,bhattacharyya1986estimates}, the assumption of rational independence for energy levels appears to not apply to long-range quantum systems with discrete energy spectrum, due to the accumulation of the energy eigenvalues at high-enegy, see \eqref{disc_spec}. This leads to the asymptotic behaviour in \eqref{rec_time_est_large_N}. Thus, the recurrence time actually decreases for long-range couplings in the thermodynamic limit and may justify the stability of the initial observable values characteristic of QSSs. 

\section{QSSs in spin systems}
\label{kit_ch}
Until now, the scenario connecting the appearance of the QSSs with the peculiarity of the spectrum in long-range systems has been presented with general arguments. In the present section, a concrete example of the occurrence of QSSs in quantum systems with discrete spectrum is presented. We consider the prototypical example of quantum criticality with long-range couplings, i.e. the long-range Ising chain. It describes quantum $1/2$-spins interacting via ferromagnetic non local couplings
\begin{align}
\label{eq:LRTFIC}
\hat{H}_{\rm LRI}=-\sum_{i=1}^{N}\sum_{r=1}^{N/2-1}t_{r}\hat{\sigma}^z_i\hat{\sigma}^z_{i+r}-h\sum_i\hat{\sigma}^x_i,
\end{align}
where $t_{r}$ is given in \eqref{hop_am}, $\hat{\sigma}^{\mu}_{i}$ is the $\mu$ component of the Pauli matrices and the indexes $i,r$ run over all sites of a one dimensional chain.

The occurrence of QSSs for the Hamiltonian in \eqref{eq:LRTFIC} has been demonstrated for a system initially prepared in an eigenstate of the transverse magnetization operator ( $\hat{m}_{x}=\sum_{i}\hat{\sigma}^x_i$), i.e. the ground state of the $h\to+\infty$ Hamiltonian, and evolved according to the $h=h_{f}=0$ Hamiltonian\,\cite{kastner2011diverging}. Within the perspective of the present work, it is rather straightforward to extend these investigations to the general $h_{i},h_{f}$ case.

It is worth noting that the Hamiltonian in \eqref{eq:LRTFIC} can be explicitly solved in the two opposite limits $\alpha\to 0,\infty$, where it represents, respectively, the fully connected Lipkin-Meshkov-Glick model\,\cite{ribeiro2007thermodynamic} and the traditional nearest-neighbour case. A convincing numerical analysis of the Hamiltonian in \eqref{eq:LRTFIC} is unfeasible in the present case, since the focus is on the thermodynamic scaling behaviour, which lies outside the available regime for standard numerical simulations. A convenient approximate representation for the Hamiltonian in \eqref{eq:LRTFIC} is the one obtained by a truncated \emph{Jordan Wigner} (JW) transformation, which reduces the problem to quadratic fermions hopping on the one-dimensional lattice\,\cite{defenu2019dynamical}\begin{align}
\hat{H}_{\rm KLR}=&\,-\sum_{i=1}^{N}\sum_{r=1}^{N/2-1}t_{r}\big(\hat{c}_i^\dagger\hat{c}_{i+r}+\hat{c}_i^\dagger\hat{c}_{i+r}^\dagger-\hat{c}_i\hat{c}_{i+r}-\hat{c}_i\hat{c}_{i+r}^\dagger\big)\nonumber\\
&-h\sum_i\big(1-2\hat{c}_i^\dagger\hat{c}_i\big).
\label{lrk_h}
\end{align}
The details of the transformation from the Hamiltonian in \eqref{eq:LRTFIC} and the one in \eqref{lrk_h} are given in Sec.\,\ref{sec:model} of the Materials and Methods.

The quadratic nature of the Hamiltonian in \eqref{lrk_h} allows its exact diagonalization in Fourier space via the Bogolyubov transformation   
\begin{align}
\label{bg_trans}
\hat{c}_{k}=u_{k}\hat{\gamma}_{k}+v^{*}_{-k}\hat{\gamma}^{\dagger}_{-k}
\end{align}
where $c_{k}=\frac{e^{i\frac{\pi}{4}}}{\sqrt{N}}\sum_{j}\hat{c}_{j}e^{-ikj}$ and the Bogolyubov angles $\theta_{k}$ are defined as
\begin{align}
\label{bg_angles}
(u_{k},v_{k})=\left(\cos\frac{\theta_{k}}{2}, \sin\frac{\theta_{k}}{2}\right),\,\,\mathrm{with}\,\,\tan\theta_{k}=\frac{\tilde{\Delta}_k}{h-\tilde{t}_k}.
\end{align}
The kinetic couplings $\tilde{t}_{k}$ are given in \eqref{ft} and the momentum space pairing term reads
\begin{align}
\label{ftp}
\tilde{\Delta}_{k}=\sum_{r=1}^{N/2-1}\sin(kr)\Delta_{r}=\frac{1}{N_{\alpha}}\sum_{r=1}^{N/2-1}\frac{\sin(kr)}{r^{\alpha}}.
\end{align}
In the $\alpha\to \infty$ the mapping becomes exact and both the fermionic and the spin models, in respectively \eqref{lrk_h} and \eqref{eq:LRTFIC}, feature two equilibrium quantum critical points at $h=h_{c}=\pm 1$, where the minimal energy gap closes at the critical momenta $k=0,\pi$ respectively. As already mentioned, effects of long-range interactions are expected to be stronger for homogeneous states. Then, our focus is on the quantum critical point appearing at $h=1$, which represents the transition point between the disordered and the ferromagnetic states in the original spin Hamiltonian. Correspondingly, the Fermi system in \eqref{lrk_h} features a transition between a topologically trivial state at $h\geq1$ and a topologically non-trivial one at $h<1$, where no local order parameter occurs\,\cite{fradkin2013field}. 

\begin{figure*}[t]
\centering
\includegraphics[width=5.9cm,height=11.4cm]{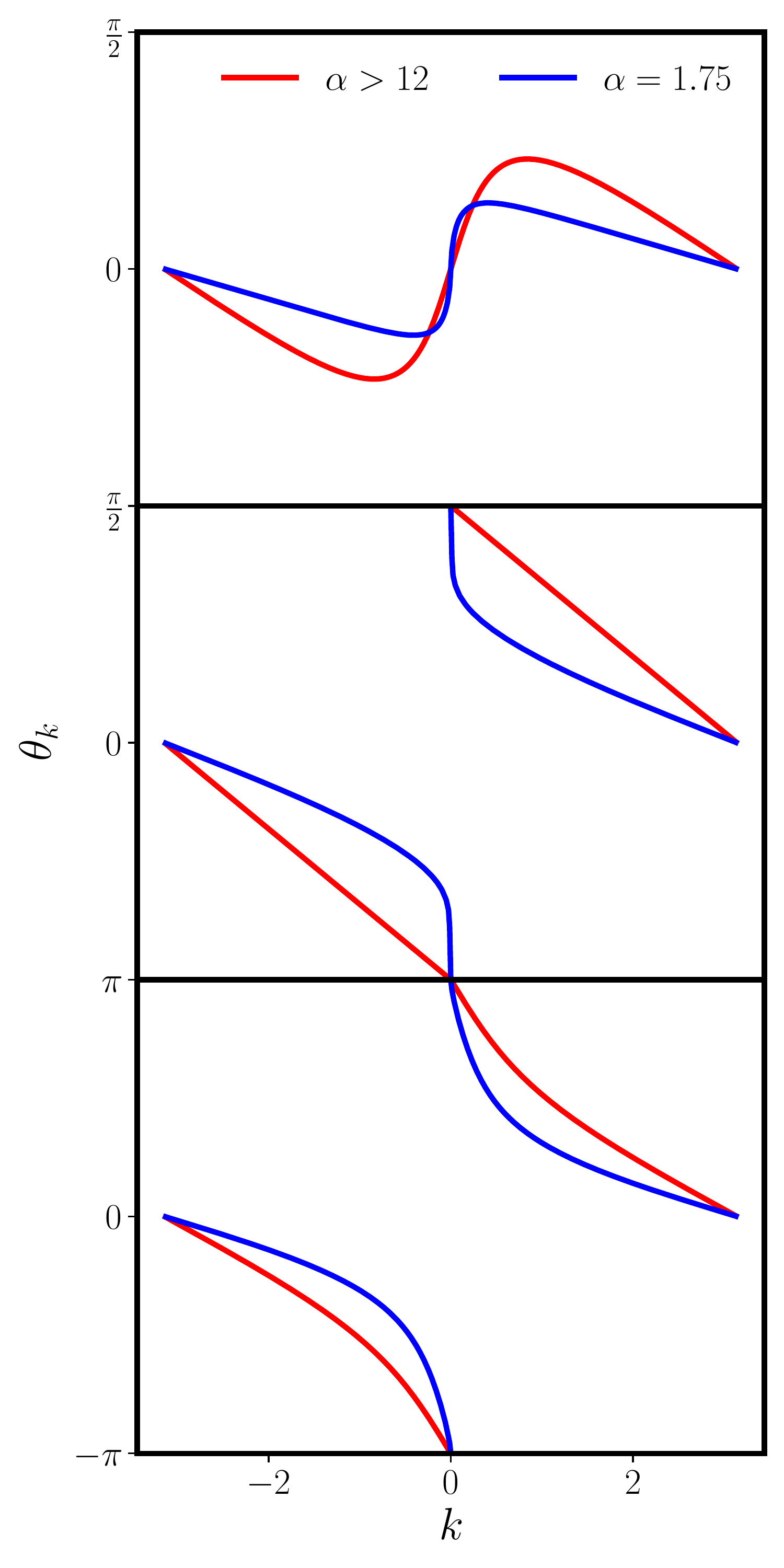}
\includegraphics[width=5.9cm,height=11.4cm]{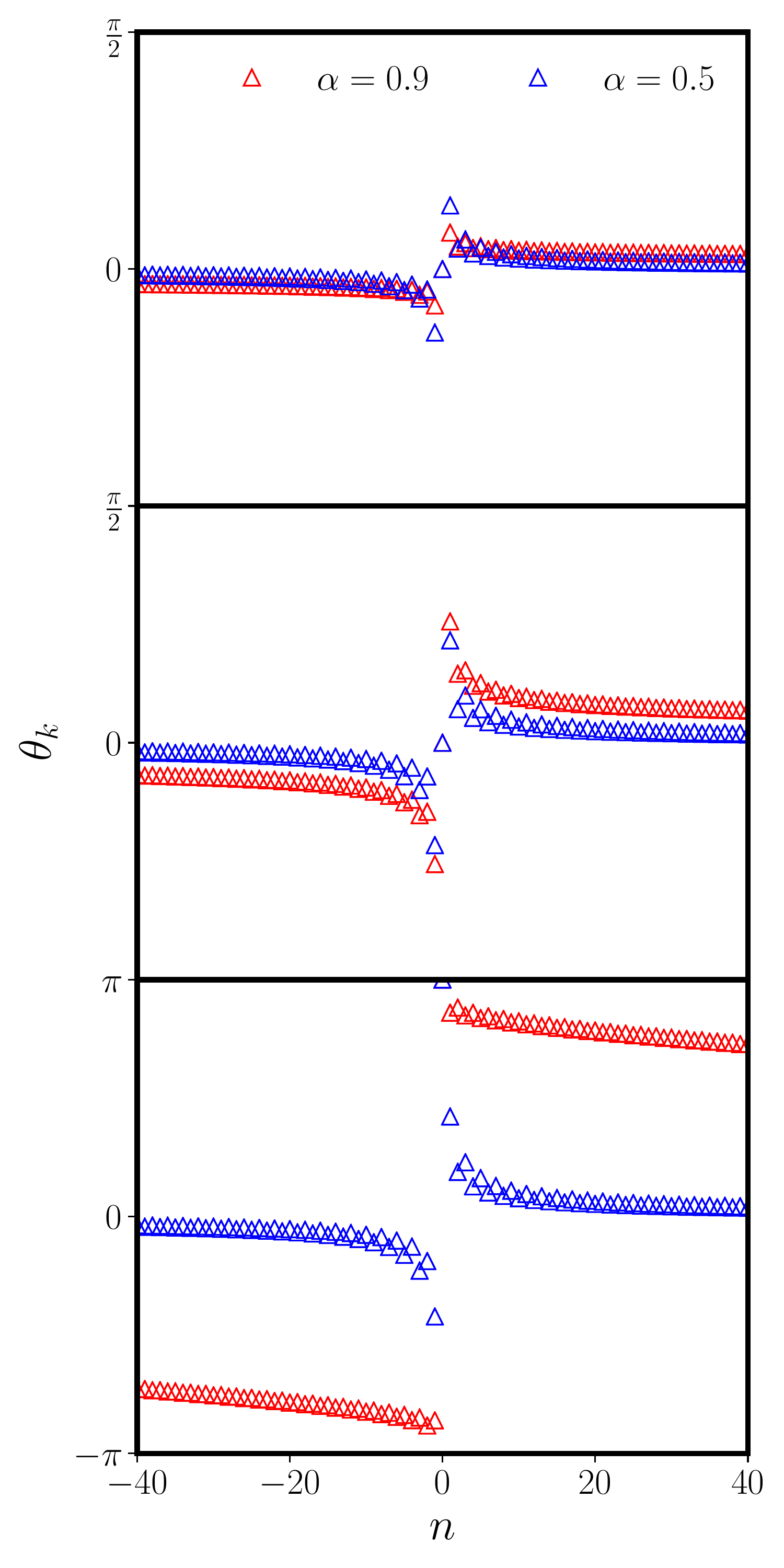}
\caption{The Bogolyubov angle of the long-range Kitaev chain as a function of the momentum, the three possible configurations $h>1$, $h=1$ and $h<1$, corresponding to trivial, critical and topological phases, are displayed in each panel from top to bottom.
In the $\alpha>1$ case (left panel), the expression for the Bogolyubov angles tends to a continuous function in the thermodynamic limit; its asymptotic behaviour in the $k\to 0$ limit is consistent with a finite winding number for $h<1$ (bottom panel on the left). Conversely, in the strong long-range  regime ($\alpha<1$) no continuous theory can be defined in the thermodynamic limit and the Bogolyubov angles are more conveniently reported as a function of the integer index $n$ (right panel). Even without a well defined notion of continuous limit, a clear distinction appears between the trivial phase at $h>1$ (upper panel on the right) and the ``topological" one at $h<1$ (lower panel on the right). It is worth noting that che characterisation of the critical phase $h=1$ is not straightforward in this case as the Bogolyubov angle $\theta_{n=0}$, which is conventionally reported as $\theta_{n=0}=0$ in the plot,  is actually indeterminate (middle panel on the right).}\label{Fig2}
\end{figure*}

For $1<\alpha<3$ the exactness of the correspondence between the fermion and spin Hamiltonians is lost. Thus, both the equilibrium and the dynamical critical properties differ\,\cite{defenu2019dynamical}. Yet, the existence of the quantum critical points is preserved and the qualitative scenario for the two systems remains quite close. As long as $\alpha>1$, the topological nature of the transition can be summarised by the small momentum limit of the Bogolyubov angles $\theta_{k}$. Indeed, for $h>1$ the denominator of the second term in \eqref{bg_angles} remains positive and $\lim_{k\to 0^{\pm}}\theta_{k}=0$, while for $h<1$ one has $\lim_{k\to 0^{\pm}}\theta_{k}=\pm\pi$, with this last discontinuity being at the origin of the finite integer winding number observed in the topological case, see left panel in Fig.\,\ref{Fig2}, upper and lower sub-panels respectively. Within this perspective, it is worth noting that for $h=h_{c}=1$ one has $\lim_{k\to 0^{\pm}}\theta_{k}=\pm\frac{\pi}{2}$, middle sub-panel in the left panel of Fig.\,\ref{Fig2}, yielding an equal superposition of electron and hole for the critical mode ($|u_{k=0}|=|v_{k=0}|=1/\sqrt{2}$), which is conventionally interpreted as the Dirac mode originating from the superposition of two Majorana edge states\,\cite{fradkin2013field}.

In the $0<\alpha<1$ regime the scenario is more complicated. Indeed, the persistence of the discrete spectrum in the thermodynamic limit, see \eqref{core_res}, does not allow to define a continuous theory and hinders the conventional definition of quantum critical point in the Kitaev chain.  Yet, the Bogolyubov angle distribution is consistent with a change of phase also in the strong long-range regime, as it is shown on the right panel of  Fig.\,\ref{Fig2}, where the different low-energy limits $n\to0$ of the two phases $h>1$ and $h<1$ are shown. We refer the reader to Sec.\,\ref{pl_coupl} of the Materials and Methods for the derivation of the momentum space hopping and pairing couplings of the Hamiltonian in \eqref{lrk_h}. Up to our knowledge, the characterisation of the $h_{c}=1$ topological phase transition in the strong long range Kitaev chain has not been discussed in the literature before, since the Kac rescaling factor was not introduced in previous studies of the kitaev Hamiltonian\,\cite{regemortel2016information}. 

In the present case, the presence of the Kac scaling factor is a direct consequence of the relation between the Hamiltonian in \eqref{lrk_h} and the original one in \eqref{eq:LRTFIC}, where the Kac scaling factor is introduced to stabilise the ferromagnetic quantum critical point. A proper investigation of the critical properties of this ``discrete topological phase" is not presented here and it will be the subject of following work. Our main concern are the peculiar equilibration properties of the model, when quenched across its equilibrium critical point at $h=1$. Hence, the abrupt modification of the Bogolyubov angles distribution for $h<1$ provides a solid enough background to consider critical those quenches, where the transverse magnetic field abruptly changes form $h_{i}\gg 1$ to $h_{f}<1$ at $t=0$, see the right panel in Fig.\,\ref{Fig2}.

Before diving into the strong long-range case, it is convenient to summarise the traditional picture for sudden quench dynamics in the nearest-neighbour Kitaev chain. The system is initially prepared in the transversally polarised initial state at $h=h_{i}\gg 1$, where $\langle \hat{m}_{x}\rangle= \langle\sum_{i}\hat{\sigma}^{x}_{i}/N\rangle\approx 1$. Then, the system is evolved according to the Hamiltonian in \eqref{lrk_h} with $h=h_{f}<1$. The explicit description of the quench dynamics solution can be found in Sec.\,\ref{sec:dynamics} of the Materials and Methods. In line with previous QSSs investigations we are going to focus on the evolution of the trasverse magnetization $\langle\hat{m}_{x}\rangle=m_{x}(t)$. The representation of the transverse magnetisation remains local also in terms of the Fermi quasi-particles, due to the relation
\begin{align}
\hat{m}_{x}=1-\frac{2}{N}\sum_{i}\hat{c}^{\dagger}_{i}\hat{c}_{i}.
\end{align}
 
\begin{figure}[tbhp]
\centering
\includegraphics[width=.8\linewidth]{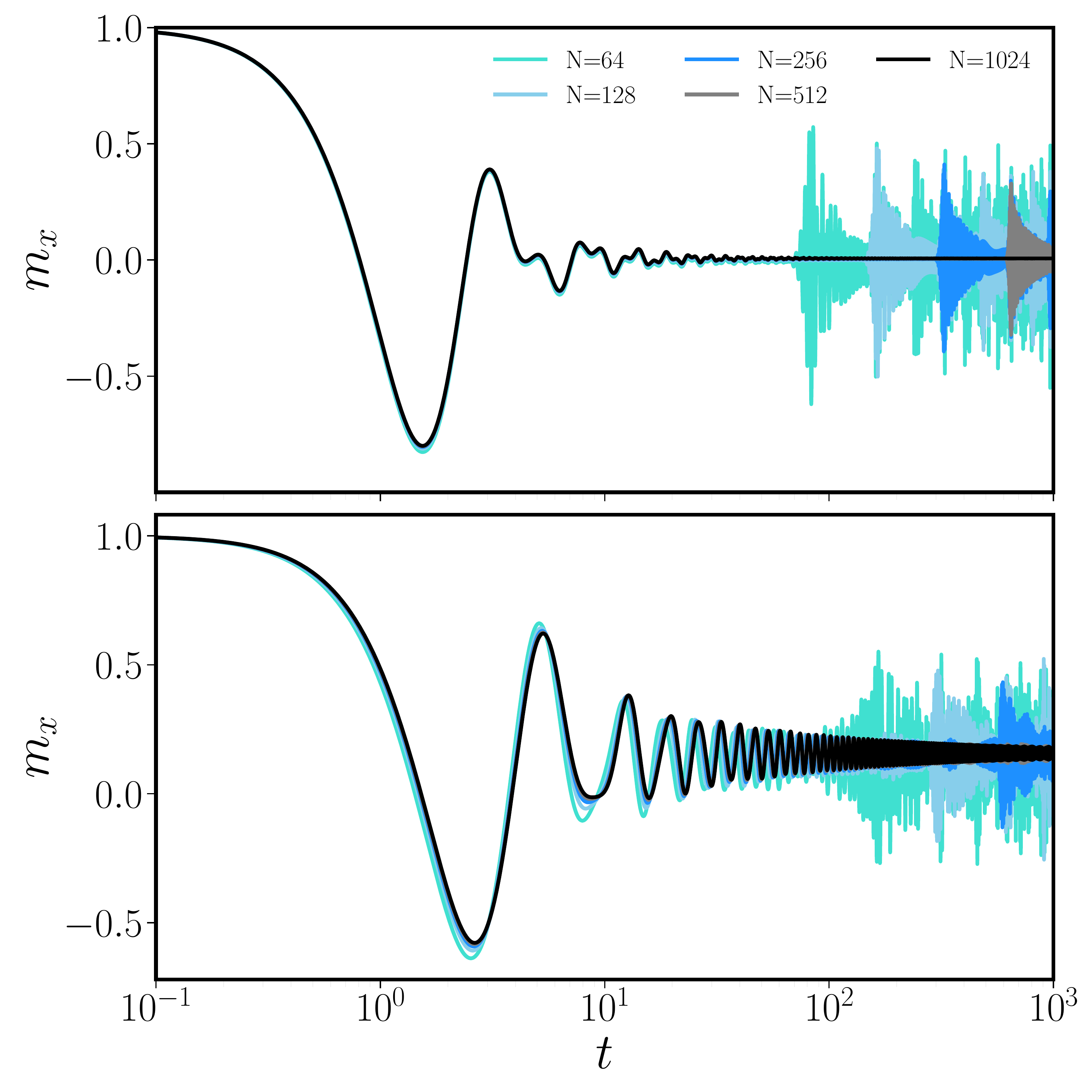}
\caption{Transverse magnetisation $m_{x}(t)$ after a quench in the long-range Ising model as represented by the dynamics of the long-range Kitaev chain in \eqref{lrk_h} in the nearest neighbour ($\alpha\gg 3$) (upper panel) and $\alpha=1.75$ (lower panel) cases. The range of values on the $y$ axis depends on the peculiar initial and final transverse field values $h_{i}$ and $h_{f}$, but the qualitative features of the equilibration remain the same for all $h_{i}>1$ and $h_{f}<1$. Each curve starts at the initial value $m_{x}=1$ and, after few oscillations, equilibrates to a constant value, which persists for a time interval, which steadily increases with the system size.}
\label{Fig0}
\end{figure}
As long as $\alpha>1$, the time evolution of the transverse magnetisation is consistent with the expectations for an integrable system. At $t=0$ the observable has its initial value and, then, rapidly equilibrates to a different constant expectation, which is maintained along the entire dynamics apart from few rapid time fluctuations appearing at the Poincar\'e recurrence times. The fluctuations  become increasingly more uncommon as the system approaches the thermodynamic limit, in agreement with the expected divergence of the recurrence times discussed in previous sections, see Fig.\,\ref{Fig0}

\begin{figure}[tbhp]
\centering
\includegraphics[width=.8\linewidth]{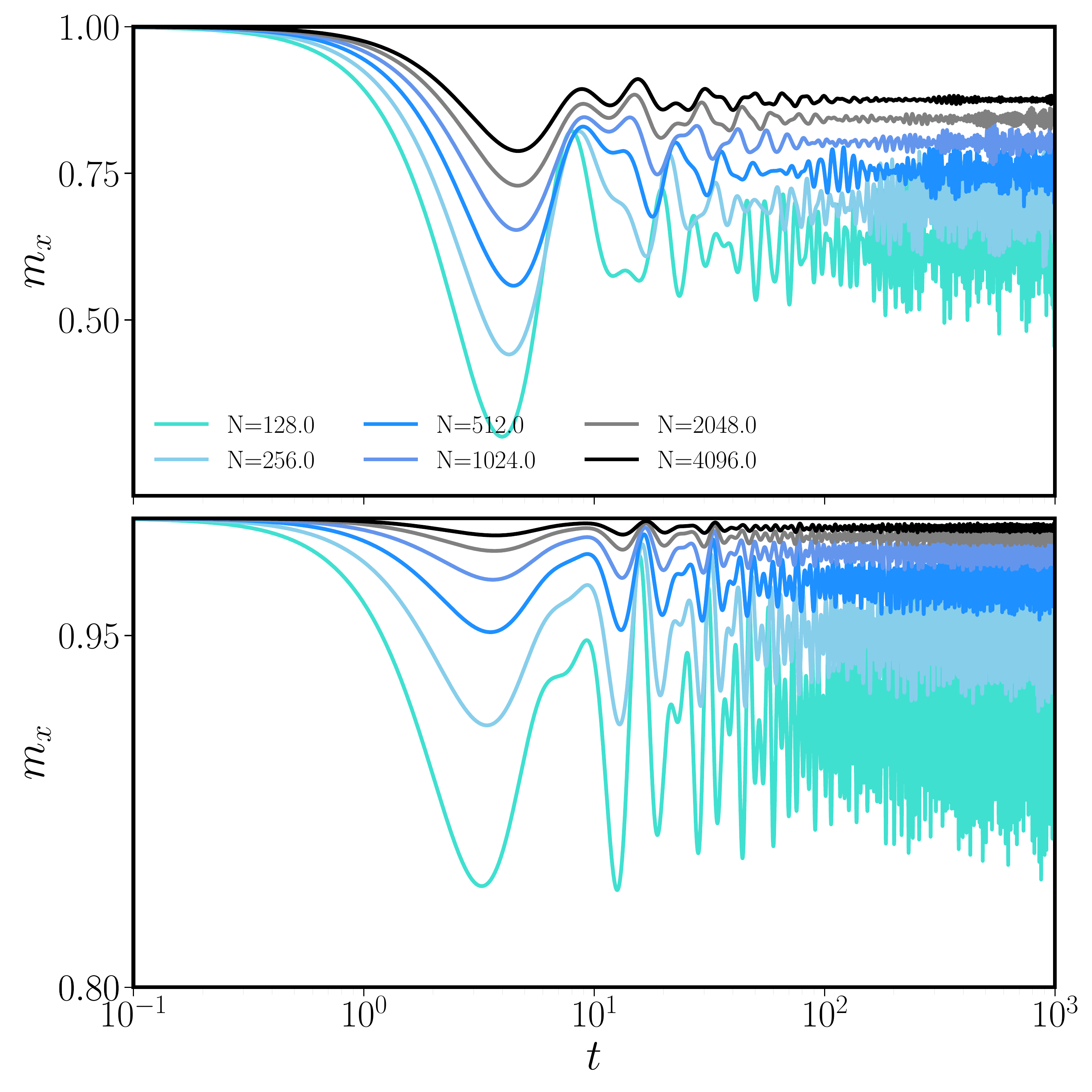}
\caption{Transverse magnetisation $m_{x}(t)$ after a quench in the long-range Ising model as represented by the dynamics of the long-range Kitaev chain in \eqref{lrk_h} in the $\alpha=0.9$ (upper panel) and $\alpha=0.4$ (lower panel) cases as a function of the size (see legend).  The range of values on the $y$ axis depends on the peculiar initial and final transverse field values $h_{i}$ and $h_{f}$, but the qualitative features of the equilibration remain the same for all $h_{i}>1$ and $h_{f}<1$. As the size of the system grows the observable large time limit changes, increasingly approaching its initial value $m_{x}=1$, and, thus, no actual equilibrium value emerges. Moreover, even if the amplitude of the oscillations tends to decrease in the $N\to\infty$ limit, the time-scale for such fluctuations is not altered by size modifications, in contrast with the conventional Poincar\'e recurrence phenomenon occurring in Fig.\,\ref{Fig0}.}
\label{Fig1}
\end{figure}
The picture is radically altered in the $\alpha<1$ case, see Fig.\,\ref{Fig1}. At intermediate system sizes the qualitative features remain similar to the $\alpha>1$ case, with the transverse magnetisation rapidly moving from its initial value $m_{x}(0)\approx 1$  to a different large-time expectation, around which it steadily oscillates. However, as the system size is increased the discrepancy with the traditional case is noticed. At larger $N$ the large-time magnetisation value tends to steadily grow, approaching the initial value $m_{x}=1$. Moreover, the time-scales of the oscillatory fluctuations are not altered by the increase in the system size, but rather manifest at almost equal time intervals at all sizes, consistently with the existence of finite recurrence times in the long-range systems at $\alpha<d$. These observations are consistent with the presence of QSSs in the long-range Kitaev chain and are analogous to the picture obtained in the long-range Ising Hamiltonian, apart qualitative differences due to the approximate relations between the two models\,\cite{kastner2011diverging}.

\section{Lack of equilibration and relation with disorder}
\label{sph_model}

The persistence of the initial observable expectation value after a quench across the critical point in long-range systems is one of the hallmarks of QSSs\,\cite{antoni1995clustering, mukamel2005breaking, campa2009statistical}. As shown in the previous sections, this phenomenon can be traced back to the spectral discreteness of thermodynamically large long-range systems and, specifically, to the accumulation of energy states in the high-energy portion of the spectrum.

The derivation in Sec.\,\ref{spec_lr_systems} implies that the  spectrum of long-range Hamiltonians is entirely pure point and, accordingly, absence of equilibration is ubiquitous in non-additive quantum systems, due to the violation of the \emph{kinematical chaos} hypothesis, see Sec.\,\ref{hth_chaos}. Indeed, the dynamics of non-additive systems features long-lived time fluctuations, whose origin can be traced back to the spectrum being dominated by the dense high-energy contribution. This phenomenon is not limited to critical quenches and observables, but it generally occurs for any dynamical protocol in purely long-range Hamiltonians.

 In this section we focus on the (lack of) equilibration appearing in long-range systems after a sudden quench of the Hamiltonian parameters within the normal phase. We show that absence of equilibration is a general feature of these systems and  compare our findings with a more traditional example in the field of disordered systems.

In this perspective, it is convenient to consider the spherical representation of the Ising Hamiltonian in \eqref{eq:LRTFIC},  which is obtained replacing the spin variables $\hat{\sigma}^{\mu}$ with the operators $\hat{s}_{i}$ and their associated momenta $\hat{p}_{i}$, obeying the harmonic oscillator commutation relations $[\hat{s}_{i},\hat{p}_{j}]=i\delta_{ij}$ and $[\hat{s}_{i},\hat{s}_{j}]=[\hat{p}_{i},\hat{p}_{j}]=0$. Some similarity between the harmonic oscillator variables and the original spins is ensured by a global constraint in the form $\langle\sum_{i}s_{i}^{2}\rangle=N/4$. The resulting Hamiltonian describes the celebrated spherical model, which is one of the workhorses in the study of disordered systems\,\cite{kosterlitz1976spherical,nieuwenhuizen1995quantum,serral2004quantum, cirano2006random,akhanjee2010spherical}, and reads
\begin{align}\label{sph_ham}
H = \frac{g}{2} \sum_i \hat{p}_i^2 + \frac{1}{2} \sum_{i,j} U_{ij} \hat{s}_i \hat{s}_j
	+ \mu \left( \sum_i \hat{s}_i^2 - \frac{N}{4} \right) \ .
\end{align} 
The parameter $g$ controls the strength of quantum fluctuations and the coupling matrix $U_{ij}$ couples all pair of sites in the linear chain. The  parameter $\mu$ plays the role of an effective chemical potential and has to be chosen in order to enforce the constraint condition, see more details in the SI Appendix. It can be shown that the dynamics of the spherical model actually corresponds to the time-dependent Hartee-Fock approximation of the Ising and $O(\mathcal{N})$ rotor models\,\cite{berges2007quantum}. In this perspective, the harmonic oscillator variables in \eqref{sph_ham} effectively represent the spin-waves of the system. 

At equilibrium, the spherical model presents a quantum critical point at $g=g_{c}$, separating the normal phase at $g>g_{c}$ from a magnetised one for $g<g_{c}$. The magnetisation is represented by a macroscopic occupation of the lowest momentum state, i.e. a condensate. The universal properties at this quantum critical point correspond to the ones of quantum $O(\mathcal{N})$ rotor models in the large $\mathcal{N}$ limit\,\cite{vojta1996quantum,sachdev2011quantum}. Given this relation, the spherical model has been employed to investigate several aspects of the out-of-equilibrium dynamics of many body systems\,\cite{sotiriadis2010quantum}, including prethermalization\,\cite{chiocchetta2017dynamical}, defect formation\,\cite{degrandi2010adiabatic} and dynamical phase transitions\,\cite{syed2021dynamical}. These studies have shown that the system's observables after a sudden quench of one of its microscopic parameters relax to an equilibrium value in the long time limit, both in the case of local interactions\,\cite{sotiriadis2010quantum,chandran2013equilibration} and in the weak long-range case $\alpha>1$\,\cite{syed2021dynamical}.

According to our picture, equilibration is not expected to occur in the non-additive regime $\alpha<1$, where the eigenvalue spectrum of the coupling matrix $U_{ij}=-\frac{J_{0}}{|i-j|^{\alpha}}$ remains discrete. A straightforward proof of this fact is obtained by studying the dynamical evolution of the system after a sudden quench of the effective chemical potential $\mu$. The system is initially prepared in the ground state with $g>g_{c}$ such that $\mu_{0}= 2\mu_{c}$. Then, at $t=0$ the constraint is suddenly lifted and the effective chemical potential switches to $\mu_{f}=\mu_{c}$. The dynamics may be characterised in terms of the evolution of the operator expectation
\begin{align}
\label{at_def}
A(t)=\left\langle\frac{4 \sum_{i}\hat{s}_{i}^{2}}{N}\right\rangle.
\end{align}
whose initial value $A(0)=1$ is fixed by the constraint. It is worth noting that the constraint is lifted for $t>0$ and, thus, the observable $A(t)$ evolves towards a long time expectation, depending on the final value of the effective chemical potential $\mu_{f}$. 

Here, we explicitly consider a quench from the normal phase $\mu>\mu_{c}$ to the vicinity of the critical point at $\mu\simeq\mu_{c}$, but the same qualitative picture is obtained for any initial and final values of the effective chemical potential. Moreover, in order to simplify both the calculations and the graphical presentation of the results, we explicitly refer to the long-time dynamics of the observable in \eqref{at_def}. However, the same exact picture on equilibration would be obtained for any relevant physical observable and, especially, for the ground-state fidelity, see the discussion in Sec.\,\ref{hth_chaos}.

  The relaxation dynamics of the observable $A(t)$ can be described in terms of its Cesaro's average  $\bar{A}=\lim_{T\to\infty}\langle A \rangle_{T}$.  In analogy with the discussion in Sec.\,\ref{hth_chaos}, see \eqref{rl_lemma} and \eqref{rage_th}, the relaxation of the observable $A(t)$ can be quantified by the parameter
\begin{align}
Q_{A}(T)=\langle\left|A(t)-\bar{A}\right|^{2}\rangle_{T}
\end{align}
which measures the width of dynamical fluctuations around the mean of $A$. In order for the observable to equilibrate the width of quantum fluctuations has to vanish in the long-time limit, leading to the general criterion
\begin{align}
\label{eq_def}
\lim_{T\to\infty}Q_{A}(T)\approx 0
\end{align}
which defines observable equilibration in closed quantum systems\,\cite{short2011equilibration,reimann2008foundation,linden2009quantum,oliveira2018equilibration}. For nearest neighbours or weak long range interactions ($\alpha>d$) the excitations spectrum of the model in \eqref{sph_ham} is (absolutely) continuous in the thermodynamic limit and, according to the Riemann--Lebesgue lemma\,\cite{hughes2008calculus}, one has $\lim_{T\to\infty}Q_{A}(T)=0$, see more details in the SI Appendix. This is not the case for strong long-range systems  ($\alpha<d$), where the complete pure point nature of the spectrum produces finite dynamical fluctuations in thermodynamically large systems $\lim_{T\to\infty}Q_{A}(T)\neq0$.
\begin{figure}[tbhp]
\centering
\includegraphics[width=1.\linewidth]{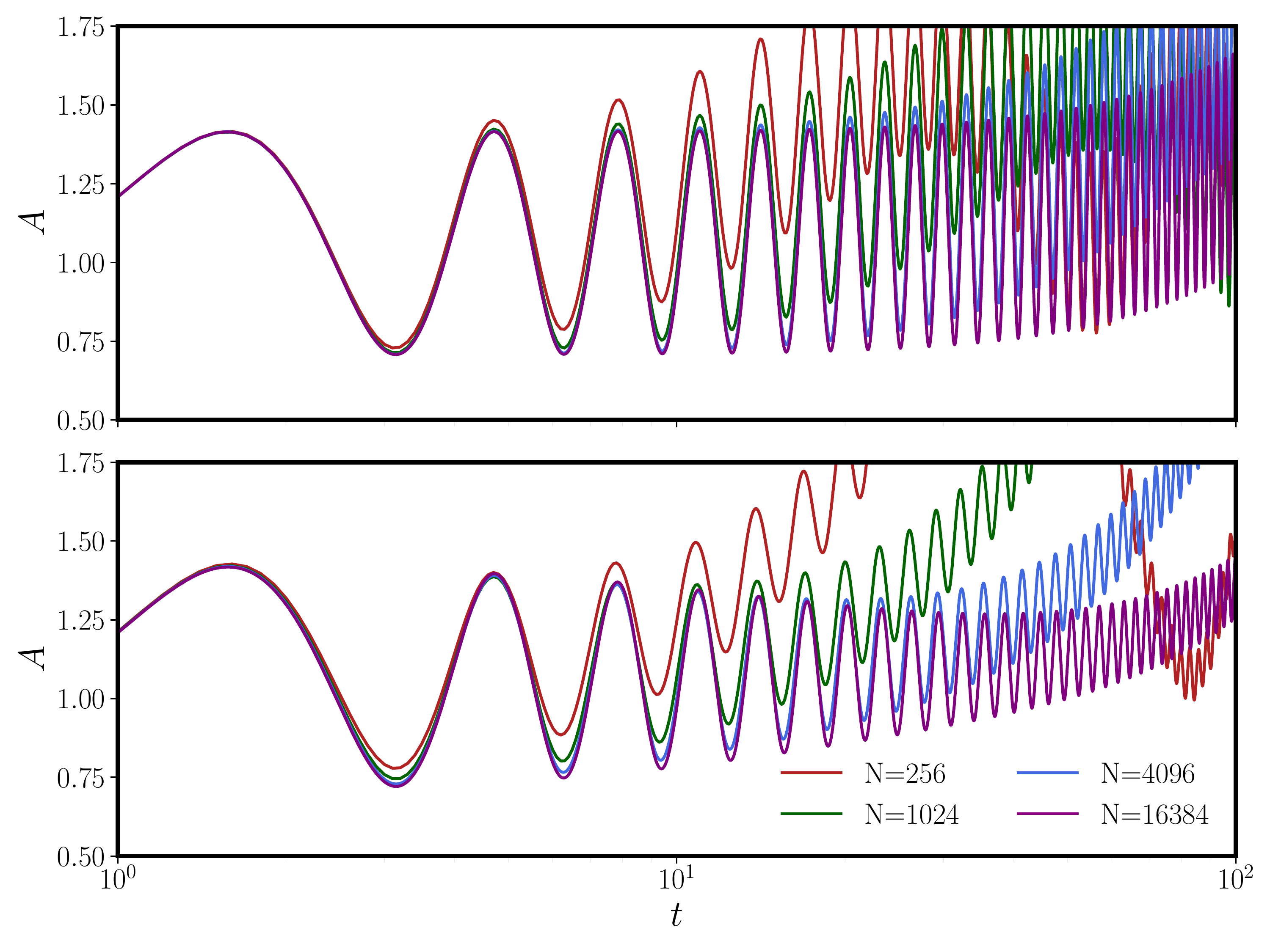}
\caption{Evolution of the observable $A(t)$ in \eqref{at_def} after a sudden quench of the effective chemical potential from the initial value $\mu_{0}\approx2\mu_{c}$ to the final value $\mu_{f}\approx \mu_{c}$ with constant $g$.  Panel (a) and (b)  represent the evolution of the quantity $A$ in a system with $\alpha=0.4,0.95$ respectively, for different system sizes $N=[2^{7},2^{9},2^{11},2^{13}]$ from top to bottom. At short times the dynamics is dominated by the high-energy contributions and the evolution is close to the one of a single harmonic oscillator, then finite size effects kick in and the oscillations of low energy modes become visible. Increasing the system size the spectral weight accumulates at high energy and the curve approaches the single oscillator limit.}
\label{Fig4}
\end{figure}

The signatures of metastability are evident in Fig.\,\ref{Fig4}. Each curve represents a different system size (largest at the bottom) for the case $\alpha=0.4,0.95$ (upper and lower panel respectively). The oscillations around the average time observable $\bar{A}$ become increasingly more regular as the system size increases, due to the dominant contribution from the dense high-energy portion of the spectrum. Then, in the thermodynamic limit the dynamical behaviour roughly approaches the one of a single harmonic oscillator, which represents the dense high-energy eigenvalue. Notice that the approach to the thermodynamic limit is slower the larger the decay rate of interactions.

\begin{figure}[tbhp]
\centering
\includegraphics[width=1.\linewidth]{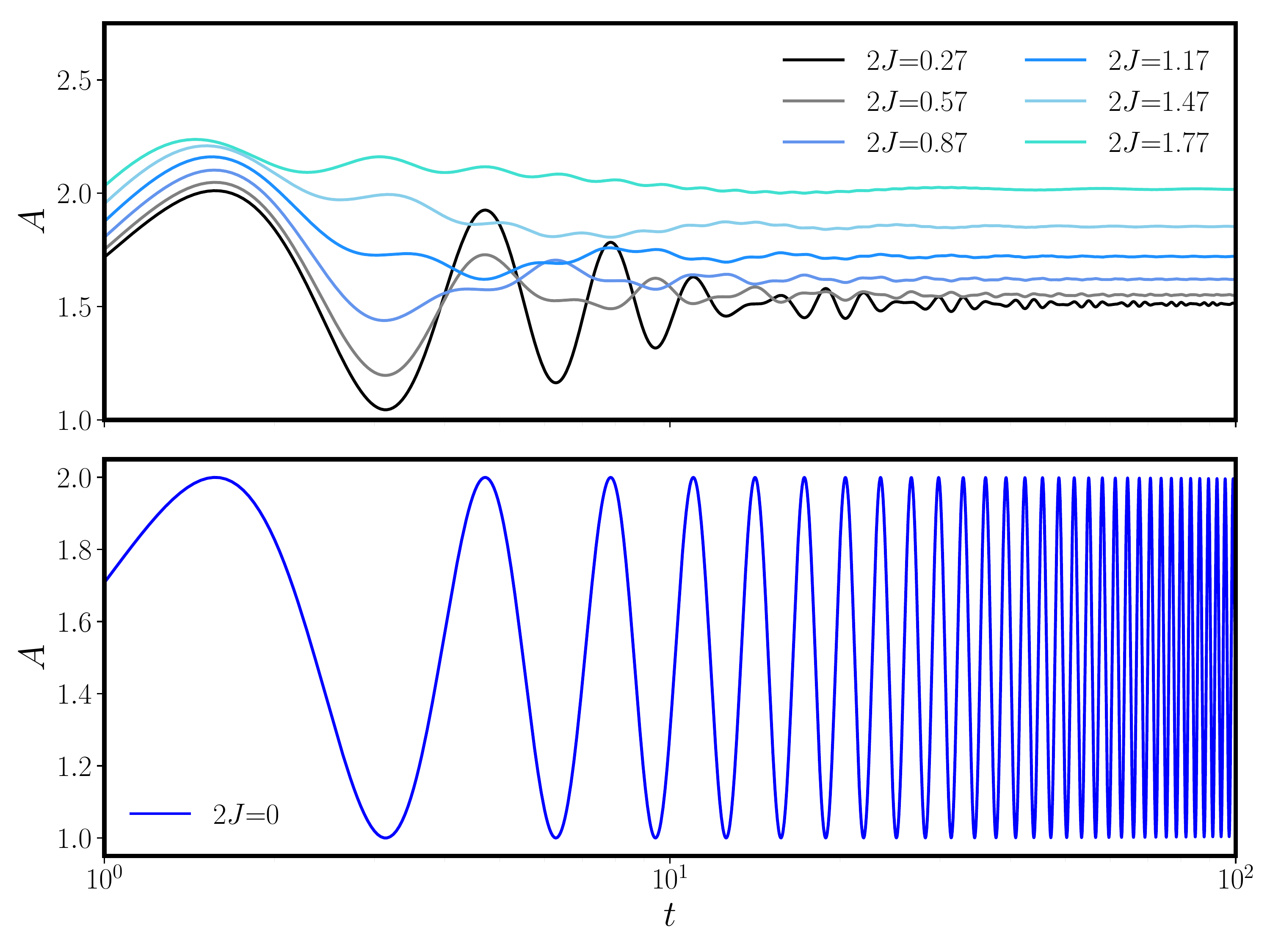}
\caption{Panel (a): evolution of the observable $A(t)$ in \eqref{at_def} after a sudden quench of the effective chemical potential from the initial value $\mu_{0}= 2\mu_{c}$ to the final value $\mu_{f}=\mu_{c}$ (with constant $g$) in the $\alpha=0$ case at finite disorder strength $2J>0$.  Each curve represents a different value of $2J\sim[0.2,2.0]$.  Dynamical fluctuations vanish in the large time limit due to the infinite Poincar\'e recurrence times caused by the continuum spectrum, see Sec.\,\ref{rec_time}.Lowering the disorder strength the dynamical evolution approaches the one of a single mode, which represents the clean (not-disordered) case and equilibration times diverge accordingly, see the SI Appendix (Fig.\,S3).  Panel (b): Same as panel (a) but in absence of disorder ($2J=0$). The dynamics exactly corresponds to a single quenched harmonic oscillator and no-equilibration occurs.}
\label{Fig5}
\end{figure}

Despite the quench protocol under study does not cross the critical point,  the scaling of Poincar\'e recurrence times approximately obeys the conjecture presented in Sec.\,\ref{rec_time}. As long as the system is finite, the dynamical evolution receives contributions from diverse frequencies in the low-energy regime and the dynamics is rather irregular, see the upper (red) curves in Fig.\,\ref{Fig4}. As the size of the system grows the spectrum becomes increasingly more dense at high energy, dispersing the contribution from the low-energy modes. As a result, the dynamical evolution steadily approaches the one of a single harmonic oscillator with energy $\omega_{\rm max}\approx\sqrt{\mu_{\rm f}}$, see the lowest (purple) curves in Fig.\,\ref{Fig4} or the thermodynamic limit solution for the $\alpha=0$ case in the lowest panel of Fig.\,\ref{Fig5}.  Thus, Poincar\'e recurrence times do not diverge in the thermodynamic limit (as they would for local systems) but rather approach their absolute minimum, which is given by the inverse of the maximum spin-wave frequency $\tau\propto (\max_{n}\omega_{n})^{-1}$.

This phenomenon is particularly evident in the limit of fully connected interactions ($\alpha\to 0$), where the spectrum separates in two distinct energy levels: a non-degenerate ground-state with energy $\varepsilon_{0}\approx -J_{0}$ and a $N-1$ degenerate excited state with energy $\varepsilon_{1}\approx 0$. Then, in absence of condensation, i.e. in the normal phase, the contribution of the ground-state can be ignored. As a consequence, the dynamics of the $\alpha=0$ spherical model corresponds to the evolution of a single harmonic oscillator with time dependent frequency, whose oscillation amplitude remains constant along the entire dynamics.

From this perspective, it is possible to provide a straightforward comparison with the case of disordered systems. The fully connected coupling matrix at $\alpha=0$ may be perturbed with the disorder contributions $u_{ij}$, which obey the Gaussian distribution $P(u_{ij})\propto \exp\left(-N\,u_{ij}^{2}/2J^{2}\right)$, where the coupling $J$ represents the disorder strength. It is worth noting that the spherical Hamiltonian in \eqref{sph_ham} with fully connected randomly distributed couplings has been one of the prototypical models for glasses, especially in its classical formulation\,\cite{kosterlitz1976spherical,cirano2006random} but also in the quantum case\,\cite{ye1993solvable,vojta1994generalized}.

 As a result of disorder the infinite degeneracy of the excited state at $\varepsilon_{1}$ of the Hamiltonian is removed and the spectrum separates into an isolated ground-state plus an high energy continuum, in analogy with the problem of a single impurity in a crystal\,\cite{kosterlitz1976spherical, edwards1976eigenvalue,dellanna2008critical}. The density of states of the high energy continuous subspace of the spectrum is distributed according to the celebrated Wigner semicircle law\,\cite{metha2004random}. Hence, the computation of the quench dynamics can be pursued directly in the thermodynamic limit, see the SI Appendix. As for the clean case, the system is prepared at equilibrium with $\mu_{0}= 2\mu_{c}$ and, then, suddenly quenched at $\mu_{f}=\mu_{c}$. The resulting evolution for the quantity $A(t)$ exponentially converges to its long time average $\bar{A}$ as a result of the Riemann--Lebesgue lemma, see Fig.\,\ref{Fig5}.

Thus, in analogy with the weak long-range case and in antithesis with the one of strong long-range interactions, the measure $Q_{A}(T)$ of dynamical fluctuations vanishes in the long time limit for any finite disorder strength and for all initial states in the normal phase. As a further proof that the continuous nature of the spectrum, resulting from disorder, causes equilibration in the system, we notice that the decay of the squared amplitude $Q_{A}(T)$ in Fig.\,\ref{Fig5} is roughly exponential $Q_{A}(T)\sim \exp(-T/\tau_{\rm eq})$, with a decay rate $\tau_{\rm eq}$ that diverges in the clean limit ($\lim_{J\to 0}\tau_{\rm eq}\to\infty$), see the numerical analysis in the SI Appendix and the result for the non-disordered case ($2J=0$) in the lower panel of Fig.\,\ref{Fig5}.

The present picture for the dynamics of the quantum spherical model with infinite range disordered couplings are analogous to the one found in the classical case with Langevin dynamics, see Chap.\,4 of Ref.\,\cite{cirano2006random}. Notably, the absence of metastable dynamics for quenches inside the normal phase or from the normal to the broken phase is common to both the classical and the quantum realms at finite disorder. It is worth noting that both the disordered and the clean systems display a non-degenerate ground-state with vanishing spectral measure, whose energy is separated from the rest of the spectrum by a finite gap. Accordingly, any initial state with a finite (macroscopic) overlap with this isolated lowest energy state, i.e. a condensate state, presents dynamical fluctuations due to the pure point nature of the low-energy spectrum. 

\section{Discussion}
\label{disc}

In the present paper, the ubiquity of long-lived metastable states  (QSSs) in the dynamical behaviour of long-range quantum systems has been connected with the impossibility of defining a continuum theory in the thermodynamic limit. Indeed, while conventional local or weak long-range systems develop a continuum spectrum at $N\to\infty$, the spectrum of strong long-range translational invariant systems ($\alpha<d$) remains discrete. Here, this picture has been explicitly proven for quadratic systems and conjectured to hold also for the more general interacting case.

In order to simplify the presentation of the results, several unnecessary assumptions have been made in the course of the derivation. Indeed, the generalisation of the result in \eqref{core_res} to the higher dimensional case or to different boundary conditions (with respect to the periodic case explicitly considered) is rather straightforward, see the SI Appendix. On the other hand, the extension of our results to the general interacting case presented below \eqref{pt_res} has to be taken with some care. In fact, long-range couplings with $\alpha<d$ are mostly expected to dominate the large scale physics and, so, stabilise the perturbation theory result at low energies. However, the addition of local quartic interactions, or weak nearest neighbour couplings, may alter the nature of the high-energy portion of the spectrum, introducing a continuous subspace. As a consequence, equilibration may be restored for initial states whose spectral weights are  concentrated on the continuum portion of the spectrum. The same mechanism has been described in details for the case of weak disorder in flat interacting systems in Sec.\,\ref{sph_model}. It is worth noting that the present discussion did not encompass the boundary case $\alpha=d=1$, which is relevant for ion crystals, a thorough analysis of a proper effective theory for these systems in the thermodynamic limit is presented in Ref.\,\cite{fishman2008structural}.

On a more general perspective, the adoption of the Kac rescaling prescription in \eqref{hop_am} and its crucial role in the derivation of \eqref{core_res} may rise doubts over the applicability of our results to actual experimental systems, where such rescaling may be difficult/impossible to implement. However, as long as the scaling factor multiplies the entire Hamiltonian, as it occurs in \eqref{lrk_h}, it only amounts to a re-definition of the time-scales of the system and does not actually alter the qualitative physics of the problem\,\cite{kastner2011diverging}. Therefore, the discreteness of the spectrum evidenced in \eqref{core_res} shall be preserved also in the unscaled case.

As discussed in the introduction, long range interactions play a prominent role in quantum computation as well as in quantum simulations\,\cite{blatt2012quantum,monroe2020programmable}. Therefore, it is important to outline the impact of the present picture on quantum annealing and state preparation in long-range systems. On the one hand, following the argument below \eqref{pt_res},  spectral discreteness improves the stability of perturbation theory and, so, it facilitates the derivation of optimal counterdiabatic dynamical protocols, capable of driving the system to a certain target state without generating excitations\,\cite{campo2013shortcuts,sels2017minimizing}. On the other hand, long-range couplings may prevent the existence of such counterdiabatic driving protocols for specific target states. Indeed, the defects generated crossing the quantum critical point of a fully-connected quantum spins system remain finite for any dynamical manipulation, including the infinitely slow drive limit\,\cite{defenu2018dynamical,defenu2020quantum}. In agreement with the slow growth of entanglement entropy noticed in these systems\,\cite{lerose2020origin}. As a consequence, the quantum accuracy threshold theorem does not apply to the case of long-range correlations between qubits\,\cite{aharonov2006fault} and the performance of quantum state preparation on these systems strongly depends on the specific case under study. 

Recently, several investigations have been performed to explore the ergodicity and thermalisation in long-range systems\,\cite{russomanno2020long,kubala2020ergodicity}. However, the present findings demonstrates that the peculiar dynamical properties of these systems may have an even more fundamental origin with respect to the non applicability of Eigenstate thermalisation hypothesis. Indeed, absence of thermalisation is a well known feature of integrable\,\cite{kinoshita2006quantum} and quasi-integrable\,\cite{prosen1998time} Hamiltonians, where the level statistics does not obey the \emph{chaotic conjecture} based on random matrix theory\,\cite{berry1981quantizing,muller2004semiclassical,bohigas1984characterization}. Yet, the result in \eqref{core_res} proves that long-range interactions evade the more basic expectation of \emph{kinematical chaos}\,\cite{lasinio1996chaotic} and present finite recurrence times up to the thermodynamic limit. This scenario advocates for a deep re-shaping of our current understanding of the basic principles of thermodynamics and many-body dynamics in quantum systems with power-law decaying couplings.

\emph{Note added:} During the completion of the present manuscript, another work appeared on the arXiv\,\cite{russomanno2020long}, where numerical results for the spectrum of the long-range Ising model at finite sizes have been presented. These results appear to be consistent with the theoretical picture presented in this work.

\matmethods{
\subsection{Model and mapping to fermions}\label{sec:model}
Our solution strategy for the dynamics of the Hamiltonian in \eqref{eq:LRTFIC} was to relate it to the dynamics of a quadratic Fermi Hamiltonian. This result has been achieved by mapping \eqref{eq:LRTFIC} onto fermions using the \emph{Jordan Wigner} (JW) transformation\,\cite{fradkin2013field}
\begin{align}
\hat{\sigma}_j^x&=1-2\hat{c}_j^\dagger\hat{c}_j,\\
\hat{\sigma}_j^y&=-\mathrm{i}\Big[\prod_{m=1}^{j-1}\big(1-2\hat{c}_m^\dagger\hat{c}_m\big)\Big]\big(\hat{c}_j-\hat{c}_j^\dagger\big),\\
\hat{\sigma}_j^z&=-\Big[\prod_{m=1}^{j-1}\big(1-2\hat{c}_m^\dagger\hat{c}_m\big)\Big]\big(\hat{c}_j+\hat{c}_j^\dagger\big),
\end{align}
where $\hat{c}_j,\hat{c}_j^\dagger$ are fermionic annihilation and creation operators, respectively, that satisfy the canonical anticommutation relations $\{\hat{c}_l,\hat{c}_j\}=0$ and $\{\hat{c}_l,\hat{c}_j^\dagger\}=\delta_{l,j}$. This renders \eqref{eq:LRTFIC} in the fermionic form
\begin{align}\nonumber
\hat{H}=&\,-\sum_{i=1}^{N}\sum_{r=1}^{N/2}t_{r}\big(\hat{c}_{i}^\dagger-\hat{c}_{i}\big)\Big[\prod_{n=i+1}^{i+r-1}\big(1-2\hat{c}_n^\dagger\hat{c}_n\big)\Big]\big(\hat{c}_{i+r}^\dagger+\hat{c}_{i+r}\big)\\\label{eq:quartic}
&-h\sum_i\big(1-2\hat{c}_i^\dagger\hat{c}_i\big).
\end{align}
The Hamiltonian in \eqref{eq:quartic} cannot be exactly solved, due to the presence of higher-than-quadratic-order terms in the fermionic operators. We employ the approximation
\begin{align}\label{eq:approx}
\prod_{n=i+1}^{i+r-1}\big(1-2\hat{c}_n^\dagger\hat{c}_n\big)=1,
\end{align}
for every $r\geq 2$, neglecting the string operators in the first line of \eqref{eq:quartic}. This \textit{truncated} JW transformation leads to the quadratic Hamiltonian in \eqref{lrk_h}, which we referred to as the long-range Kitaev chain. The model in the $\alpha\to\infty$ limit reduces to the paradigmatic Kitaev chain at equal nearest-neighbours hopping and pairing strengths, which exactly represents the problem of the nearest-neighbour Ising model. 

The Hamiltonian~\eqref{lrk_h} is translation invariant and is thus more conveniently represented in Fourier space as
\begin{align}
\label{h_klr}
\hat{H}=\sum_{k}^{\rm B.\,z.}\left[(\hat{c}^{\dagger}_{k}\hat{c}_{k}
-\hat{c}_{-k}\hat{c}^{\dagger}_{-k})\varepsilon_{k}+(\hat{c}^{\dagger}_{k}\hat{c}^{\dagger}_{-k}+\hat{c}_{-k}\hat{c}_{k})\Delta_{k}\right],
\end{align}
where $\varepsilon_{k}=h-\tilde{t}_{k}$ and $\tilde{\Delta}_{k}$ is given in \eqref{ftp} . The diagonalization of the Hamiltonian in \eqref{h_klr} can be readily performed by the Bogolyubov transformation in \eqref{bg_angles}. 
\subsection{Quench Dynamics}
\label{sec:dynamics}
In Figs.\,\ref{Fig0} and\,\ref{Fig1} the dynamical evolution of the system has been studied after a quench from the ground state of the Hamiltonian in \eqref{lrk_h} far in the transverse magnetised phase, i.e. $h_{i}\gg 1$, with $\theta_{k}\approx 0$ independently on $k$. Then, the system is evolved according to the final Hamiltonian with $h=h_{f}<1$ ($h_{f}=0.4$ in the figures, but the same qualitative picture has been verified for several other values of $h_{i}$ and $h_{f}$).

The dynamics of the system has been obtained via the Heisenberg equations of motion for the original creation and annihilation operators, $i\partial_{t}\hat{c}_{k}=[\hat{c}_{k},\hat{H}]$. Latter equations can be cast into a matrix evolution for the Bogolyubov coefficients,
\begin{align}
i\partial_{t}\begin{pmatrix}u_{k}\\v_{k}\end{pmatrix}=2\begin{pmatrix}\varepsilon_{k} & \Delta_{k}\\
\Delta_{k} & -\varepsilon_{k}
\end{pmatrix}\begin{pmatrix}u_{k}\\v_{k}\end{pmatrix}.
\end{align}
For a time independent Hamiltonian the solution is simply obtained diagonalising the time evolution via the matrix
\begin{align}
U=\begin{pmatrix}\cos\frac{\theta_{k}}{2} &\sin\frac{\theta_{k}}{2}\\
-\sin\frac{\theta_{k}}{2} & \cos\frac{\theta_{k}}{2}
\end{pmatrix}
\end{align}
which is a unitary matrix. The unitary transformation $U$ brings $H_{k}$ to diagonal form with eigenvalues $\pm\omega_{k}$, it follows that the coefficients defined as
\begin{align}
\begin{pmatrix}s^{+}_{k}\\s^{-}_{k}\end{pmatrix}=U\begin{pmatrix}u_{k}\\v_{k}\end{pmatrix}\end{align}
evolve as simple plane waves $s_{k}^{\pm}(t)=s(0)e^{\mp i\omega_{k}t}$. Then, we can deduce the evolution operator for the Bogolyubov coefficients 
\begin{align}
\begin{pmatrix}u_{k}(t)\\v_{k}(t)\end{pmatrix}=E(t)\begin{pmatrix}u_{k}(0)\\v_{k}(0)\end{pmatrix}
\end{align}
with
\begin{align}
& E(t)=\begin{pmatrix}
\cos(\omega_{k}t)-i\cos(\theta_{k})\sin(\omega_{k}t) &i\sin\theta_{k}\sin(\omega_{k}t)\\
i\sin\theta_{k}\sin(\omega_{k}t)& \cos(\omega_{k}t)+i\cos\theta_{k}\sin(\omega_{k}t)
\end{pmatrix}
\end{align}
which yielded the numerical curves shown in Figs.\,\ref{Fig0} and\,\ref{Fig1}.

\subsection{Power-law couplings}
\label{pl_coupl}
Let us re-consider the Fourier transform of the long-range couplings in the Hamiltonian
in \eqref{lrk_h}. For $\alpha>1$ the normalisation only introduces a finite coefficient $N_{\alpha}=\zeta(\alpha)$ in the thermodynamic limit, which fixes the equilibrium critical point of the model at $h_{c}^e=\pm1$ irrespective of the value of $\alpha$. 

Thus, one can directly consider the $N\to\infty$ limit of the Fourier transform of the hopping and pairing couplings in \eqref{lrk_h} 
	\begin{align}
	\tilde{t}_k&=\frac{1}{\zeta(\alpha)}\sum_{r=1}^{\infty}\frac{\cos(kr)}{r^{\alpha}}=\frac{\mathrm{Re[\,Li}\left(e^{ik}\right)]}{2\zeta(\alpha)},\label{kin_tf}\\
	\tilde{\Delta}_k&=\frac{1}{\zeta(\alpha)}\sum_{r=1}^{\infty}\frac{\sin(kr)}{r^{\alpha}}=\frac{\mathrm{Im[\,Li}\left(e^{ik}\right)]}{2\zeta(\alpha)},\label{par_tf}
	\end{align}
	where the $\zeta(\alpha)$ normalization forces  the Fourier coefficient $\tilde{t}_{k=0}$  to be $1$ and the momentum now takes continuous values $k\in[-\pi,\pi]$. 
	
In the $\alpha<1$ case the $k=0$ term in the hopping amplitudes diverges in the thermodynamic limit and so does the Kac's scaling term according to \eqref{k_norm}.	 Therefore, the analytical computation of the summations in \eqref{ft} and \eqref{ftp} in the $N\to\infty$ limit requires particular care. One can rewrite the momentum space couplings as 
\begin{align}
	\tilde{t}_k&=\frac{c_{\alpha}}{N}\sum_{r=1}^{\frac{N}{2}-1}\frac{\cos(kr)}{\left(\frac{r}{N}\right)^{\alpha}}=\frac{c_{\alpha}}{N}\sum_{r=1}^{\frac{N}{2}-1}\frac{\cos(2\pi m r/N)}{\left(\frac{r}{N}\right)^{\alpha}}\label{kin_tf_app}\\
	\tilde{\Delta}_k&=\frac{c_{\alpha}}{N}\sum_{r=1}^{\frac{N}{2}-1}\frac{\sin(kr)}{\left(\frac{r}{N}\right)^{\alpha}}=\frac{c_{\alpha}}{N}\sum_{r=1}^{\frac{N}{2}-1}\frac{\sin(2\pi m r/N)}{\left(\frac{r}{N}\right)^{\alpha}},\label{par_tf_app}
	\end{align}
where we employed the explicit form for the lattice momenta with periodic boundary conditions
\begin{align}
k\equiv \frac{2\pi m}{N}
\end{align} 
where $m\in \mathbb{Z}$. Using the Right-hand Rectangular Approximation Method, the summations can be approximated with the integrals
\begin{align}
\label{t_m}
	\tilde{t}_m&= c_{\alpha}\int_{0}^{\frac{1}{2}}\frac{\cos\left(2\pi m x\right)}{x^{\alpha}}dx\\
\label{d_m}
	\tilde{\Delta}_m&=c_{\alpha}\int_{0}^{\frac{1}{2}}\frac{\sin\left(2\pi m x\right)}{x^{\alpha}}dx.
	\end{align}
The above formulas become exact in the $N\to\infty$ limit. 
In order to obtain the result in \eqref{t_m} and \eqref{d_m}, we have taken the continuous limit of the spatial variable $x\equiv r/N$,  leaving the integration boundaries over $x$ finite. The difference with the ‘‘traditional" thermodynamic limit procedure is striking, as in the present case the momentum space variable $k$, cannot be considered continuous anymore, but it remains discrete and labeled by the integer values $m$. Inserting the results in \eqref{t_m} and \eqref{d_m} into the expression for the Bogolyubov angles in \eqref{bg_angles}, one finds the discrete functions shown in the right panel of Fig.\,\ref{Fig2} for $h\gtrsim20$, $h=1$ and $h=0.4$, respectively from top to bottom. For comparison the Bogolyubov angles obtained by the continuous spectra of the Kitaev chain at $\alpha\approx 12$ and $\alpha=1.75$ are shown on the left panel.

}

\showmatmethods{} 

\acknow{Useful discussions with T. Enss, G. Folena, G. Gori, G. M. Graf, M. Kastner, G. Morigi, S. Ruffo, M. Salmhofer and A. Trombettoni are gratefully acknowledged. This work is supported by the Deutsche Forschungsgemeinschaft (DFG, German Research Foundation) under Germany’s Excellence Strategy EXC2181/1-390900948 (the Heidelberg STRUCTURES Excellence Cluster).}

\showacknow{} 
\subsection*{References}
\appendix
\section{Discrete spectrum of long-range systems in $d$-dimension}
In the following, we are going to extend the arguments of \emph{Spectrum of Long-Range Systems} to the $d$ dimensional case. Let us consider the $d$-dimensional version of the Hamiltonian in Eq.\,(7)
\begin{align}
\label{h1}
\hat{H}=-\sum_{|\vec{i}-\vec{j}|=1}^{L/2-1}t_{|\vec{i}-\vec{j}|}(\hat{a}^{\dagger}_{i}\hat{a}_{j}+h.c.)+\mu\sum_{i=1}^{N}\hat{a}_{i}^{\dagger}\hat{a}_{i}+\hat{H}_{\rm int},\end{align}
where the $\hat{a}_{i}^{\dagger}(\hat{a}_{i})$ symbols represent the creation(annihilation) operators of quantum particles on the sites at position $\vec{i}$ of a $d$-dimensional lattice. Neither the nature (bosonic or fermionic) of the particles nor the specific shape of the interaction Hamiltonian $\hat{H}_{\rm int}$ are crucial to our arguments. 

The long-range hopping amplitudes take the form,
\begin{align}
\label{hop_am}
t_{\vec{R}}=\frac{1}{N_{\alpha}}\frac{1}{|\vec{R}|^{\alpha}},
\end{align}
where $\vec{R}$ is the distance between the sites at positions $\vec{i}$ and $\vec{j}$ in the $d$-dimensional space. The factor $N_{\alpha}=\sum_{\vec{R}}|\vec{R}|^{-\alpha}$ has to be introduced in order to yield an extensive internal energy for the system\,\cite{kac1963van}. In the thermodynamic limit the normalisation factor scales according to
\begin{align}
\label{k_norm}
N_{\alpha}^{-1}\propto\begin{cases}
L^{\alpha-d}&\quad\mathrm{if}\,\,\alpha<d\\
1/\log(L)&\quad\mathrm{if}\,\,\alpha=d\\
1&\quad\mathrm{if}\,\,\alpha>d.
\end{cases}
\end{align}
 where $L$ is the linear size of the systems. Then, in the large $L$ limit for $\alpha<d$, the Fourier transform of the kinetic couplings reads
\begin{align}
\label{multid_core}
t_{\vec{k}}\approx\frac{c_{\alpha}'}{L^{d}}\sum_{\vec{R}}\frac{f(k_{1}\ell_{1},\cdots,k_{d}\ell_{d})}{(|\vec{R}|/L)^{\alpha}}\propto\int\frac{f(2\pi n_{1}s_{1},\cdots,2\pi n_{d}s_{d})}{\sqrt{s_{1}^{2}+\cdots+s_{d}^{2}}^{\alpha}}d^{d}s,
\end{align}
where $\ell_{\mu}$, with $\mu=\{1,\cdots,d\}$, are the discrete coordinates of the sites of the $d$-dimensional lattice, $k_{\mu}=2\pi n_{\mu}/L$ are the corresponding momentum vector components and $f(\{k_{\mu}\ell_{\mu}\})$ is a lattice dependent function. Then, the Riemann summation formula can be applied independently to each lattice direction $\ell_{\mu}/L\to s_{\mu}$ and the continuous $d$-dimensional integral on the right hand side. of \eqref{multid_core} is obtained. The net result is analogous to the $d=1$ case, since for each set of $d$ integers $\{n_{\mu}\}$, which label the momentum, one obtains a different result for the kinetic coupling $t_{\vec{k}}$. It is worth noting that depending on the peculiar shape of the function $f(\{k_{\mu}\ell_{\mu}\})$ (and, then, on the specific lattice symmetries) a finite number of degeneracies may occur. As long as these degeneracies are finite, they do not hinder the application of the perturbative argument below Eq.\,(15) in the main text, which also applies to $d>1$.\section*{Quantum spherical model with infinite range couplings}
\subsection*{Static}
In order to investigate the interplay between long-range couplings and disorder it is convenient to focus on the quantum extension of the Spherical model introduced by Berlin and Kac\,\cite{berlin1952spherical}. The model consists of a set  of quantum harmonic oscillators, whose average position is constrained\,\cite{vojta1996quantum}. 
The interplay between the zero point quantum fluctuations of the problem and the constraint gives rise to a quantum critical point, whose salient properties depend on the spatial dimension and the interaction shape. Our interest in this problem is justified by the fact that its free energy exactly corresponds to the one of quantum $O(\mathcal{N})$-symmetric rotors in the $\mathcal{N} \rightarrow \infty$ limit\,\cite{sachdev2011quantum,vojta1996quantum}.
In presence of in-homogeneity, such as disorder, the exact correspondence between the spherical model and the large-$\mathcal{N}$ limit of $O(\mathcal{N})$ does not hold, but universal scaling both in and out of equilibrium are expected to remain the same\,\cite{joyce1966spherical,defenu2015fixed,defenu2017criticality,syed2021dynamical}. 

Therefore,  the quantum spherical model represents a privileged tool to investigate the qualitative features caused by long-range interactions in quantum dynamics. Its Hamiltonian reads
\begin{align}\label{eq:ham_eq}
H = \frac{g}{2} \sum_i \hat{p}_i^2 + \frac{1}{2} \sum_{i,j} U_{ij} \hat{s}_i \hat{s}_j
	+ \mu \left( \sum_i \hat{s}_i^2 - \frac{N}{4} \right) \ ,
\end{align}
where the $\hat{s}_i$ and $\hat{p}_i$ are canonically conjugate
hermitian operators on the one-dimensional lattice, with harmonic oscillator commutation relations 
$\comm{\hat{s}_i}{\hat{p}_j} = \ir \delta_{i,j}$ (with $\hbar=1$). The strength of quantum fluctuations depends on the magnitude of the coupling $g$, which tends to delocalize the exctiations in the system. For $g=0$, the Hamiltonian in  \eqref{eq:ham_eq} represents the zero temperature limit of the classical spherical model\,\cite{joyce1966spherical}. The Lagrange multiplier $\mu$ has to be chosen in order to satisfy the spherical constraint 
$\ev{\sum_i \hat{s}_i^2} = \frac{N}{4}$.  

The spherical model Hamiltonian in \eqref{eq:ham_eq} is more conveniently represented in the diagonal basis of the coupling matrix $U_{ij}$, yielding
\begin{align}\label{eq:qho}
	H = \frac{g}{2} \sum_\lambda \hat{p}_\lambda^{2} + 
		\frac{1}{2g} \sum_\lambda \omega_\lambda^2 \hat{s}_\lambda^{2} 
\end{align}
with frequencies $\omega_\lambda^2 = 2g(\mu + U_\lambda / 2)$, where $U_{\lambda}$ are the eigenvalues of the matrix $U_{ij}$, which are labeled by the index $\lambda$. The operators $\hat{p}_{\lambda}$\,($\hat{s}_{\lambda}$) are projection of the real space vector operators $(\hat{p}_{1},\cdots,\hat{p}_{N})$ onto the eigenvectors associated with the eigenvalues $U_{\lambda}$. It is worth noting that in the translational invariant case $U_{ij}\equiv U_{|i-j|}$ the diagonal basis of $U_{ij}$ is obtained by the Fourier transform of the vector operator $(\hat{p}_{1},\cdots, \hat{p}_{N})$.

It is straightforward to apply the well-known definitions of the harmonic oscillator creation ($a^{\dagger}$) and annihilation ($a$) operators and obtain
\begin{align}
	H = \sum_\lambda \omega_\lambda \left( \ad_\lambda \hat{a}_\lambda + \frac{1}{2} \right).
\end{align}
Accordingly, the spherical constraint has to be evaluated in the harmonic oscillators' ground state $\ket{0} = \prod_\lambda \ket{\lambda,0}$, yielding
\begin{align}\label{eq:sph}
	\frac{4}{N}\ev{\sum_i \hat{s}_i^2}= \frac{\sqrt{g}}{N} \sum_\lambda \frac{2}{\sqrt{2\mu+U_{\lambda}}}=1.
\end{align}
In order to deal both with the clean and disordered case it is convenient to rewrite the constraint equation in terms of the density of states $\rho(\varepsilon)=\sum_{\lambda}\delta(\varepsilon-U_{\lambda})/N$, which reduces the constrain condition to
\begin{align}
\label{constraint_dos}
    \int  \frac{2\rho(\varepsilon)d\varepsilon}{\sqrt{2\mu+\varepsilon}}=\frac{1}{\sqrt{g}}
\end{align}
Following \eqref{constraint_dos}, the existence of the quantum phase transition can be inferred along the same lines as in the theory of Bose-Einstein condensation\,\cite{sachdev2011quantum}.  Indeed, lowering the strength of quantum fluctuations reduces the value of the Lagrange multiplier $\mu$ until the value $\muc=-\min_{\lambda}{U_{\lambda}}/2=-U_{\lambda^{*}}/2$,  where the integral on the right hand side of \eqref{constraint_dos} attains its maximum. As long as this maximum is infinite, the constraint condition in \eqref{constraint_dos} can be satisfied up to the $g\to 0$ limit. However, if the integral on the left hand side of \eqref{constraint_dos} remains finite at $\mu=\muc$, it exists a coupling strength value $\gc$, given by
\begin{align}
\label{gc}
    \frac{1}{\sqrt{\gc}}=\int  \frac{2\rho(\varepsilon)d\varepsilon}{\sqrt{2\muc+\varepsilon}},
\end{align}
such that for $g<\gc$
\eqref{constraint_dos} cannot be applied in its present form. Then, one has to allow for a finite macroscopic population of the lowest lying eigenstate $\langle \hat{s}_{\lambda^{*}}\rangle\neq 0$, which may be interpreted as a finite magnetization/condensation\,\cite{vojta1996quantum}.  

Therefore, the existence of the critical point at $\gc$ is solely determined by the low-energy behaviour of the density of states. As in the main text, we consider long-range couplings of the form 
$U_{ij}=-\frac{J_{0}}{N_{\alpha}}\frac{1}{|i-j|^{\alpha}}$, with the Kac normalization $N_{\alpha}=\sum_{r=1}r^{-\alpha}$\,\cite{kac1963van}. As long as $\alpha>1$ the density of states converges to a continuous function in the thermodynamic limit and the low energy scaling of the density of states reads $\rho(\varepsilon)\approx (\varepsilon-\mu_{c})^{\frac{2-\alpha}{\alpha-1}}$. Given the aforementioned low-energy scaling of the density of states the critical coupling $\gc$ is only finite for $\alpha<3$. 

Then, the transition persists for any $\alpha<3$ and, especially, for $\alpha<1$, where the density of states does not converge to a continuous function in the thermodynamic limit, but rather remains a set of discrete energy levels. In this perspective, it is particularly instructive to consider the simplest $\alpha=0$ case, i.e. $U_{ij}=-J_{0}/N\,\,\forall\,\,i,j$. With this definition the interaction energy has two possible eigenvalues in a finite chain of $N$ sites, a non-degenerate ground state $U_{\lambda=0}=-J_{0}(1-1/N)=\varepsilon_{0}$ and a $N-1$ degenerate excited state with energy $U_{\lambda>0}=J_{0}/N=\varepsilon_{1}$. As a consequence, the density of states becomes
\begin{align}
\label{dos_clean_case}
    \rho(\varepsilon)=\frac{\delta(\varepsilon-\varepsilon_{0})}{N}+\frac{N-1}{N}\delta(\varepsilon-\varepsilon_{1}),
\end{align}
so that in the thermodynamic limit the spectral weight accumulates at large energy ($\varepsilon_{1}>\varepsilon_{0}$) and the contribution of the ground-state becomes negligible. Inserting \eqref{dos_clean_case} into \eqref{gc} immediately yields the result $g_{c}=J_{0}/4$ as expected for the mean-field spherical model\,\cite{dellanna2008critical}. A similar phenomenon occurs for any $0<\alpha<1$. There, at finite sizes, the eigenstates $\varepsilon_{n}>\varepsilon_{0}$ are non degenerate, but, approaching the thermodynamic limit, they accumulate at high energy, where the level splitting vanishes. Hence, the thermodynamic limit behaviour in the entire range $0<\alpha<d$ resembles the one of the flat interactions $\alpha=0$ case, since the sum in \eqref{eq:sph} is dominated by the dense (and degenerate) high energy contribution. 

\begin{figure}[ht!]	
\centering
\includegraphics[width=1\linewidth]{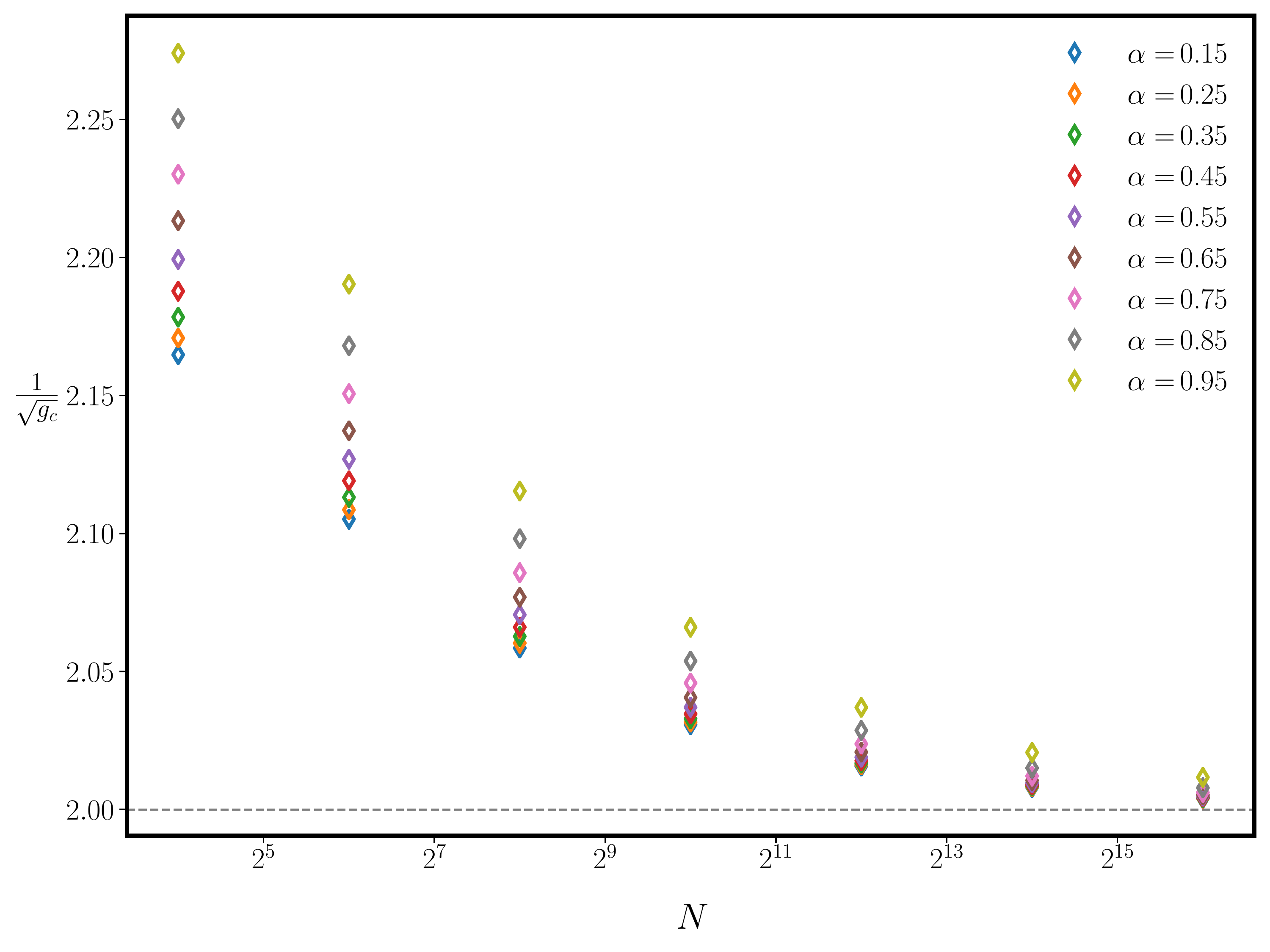}
	\caption{The finite size scaling of the critical coupling $\gc$ for several values of $\alpha\in[0.15,0.95]$ from top to bottom. All results converge to the same limit $\gc=1/4$ (grey dashed line) irrespectively on the $\alpha$ value (we set $J_{0}=1$ in the plot). This proves that the integral in \eqref{eq:sph} is always dominated by the dense accumulation point at high energy, whose eigenvalue is $\max_{\lambda}U_{\lambda}=0$ independently on $\alpha$.  }
\label{Fig0app}
\end{figure}
A first indication of this phenomenon is already visible in the finite size scaling of the critical coupling obtained via \eqref{gc}, which converges to the ($\alpha=0$) flat interactions result $\gc=J_{0}/4$ irrespectively of the $\alpha$ value, see Fig.\,\ref{Fig0app}. The dynamical counterpart of this phenomenon is discussed in the following section and in \emph{Lack of Equilibration and Relation with Disorder}, see the main text. In particular, Fig.\,4 of the main text displays the dynamical evolution of the observable $\langle \sum_{i}\hat{s}_{i}^{2}\rangle/N$, which is shown to converge to the $\alpha=0$ result, see the lower panel of Fig.\,5.

\begin{figure}[ht!]	
\centering
\includegraphics[width=1\linewidth]{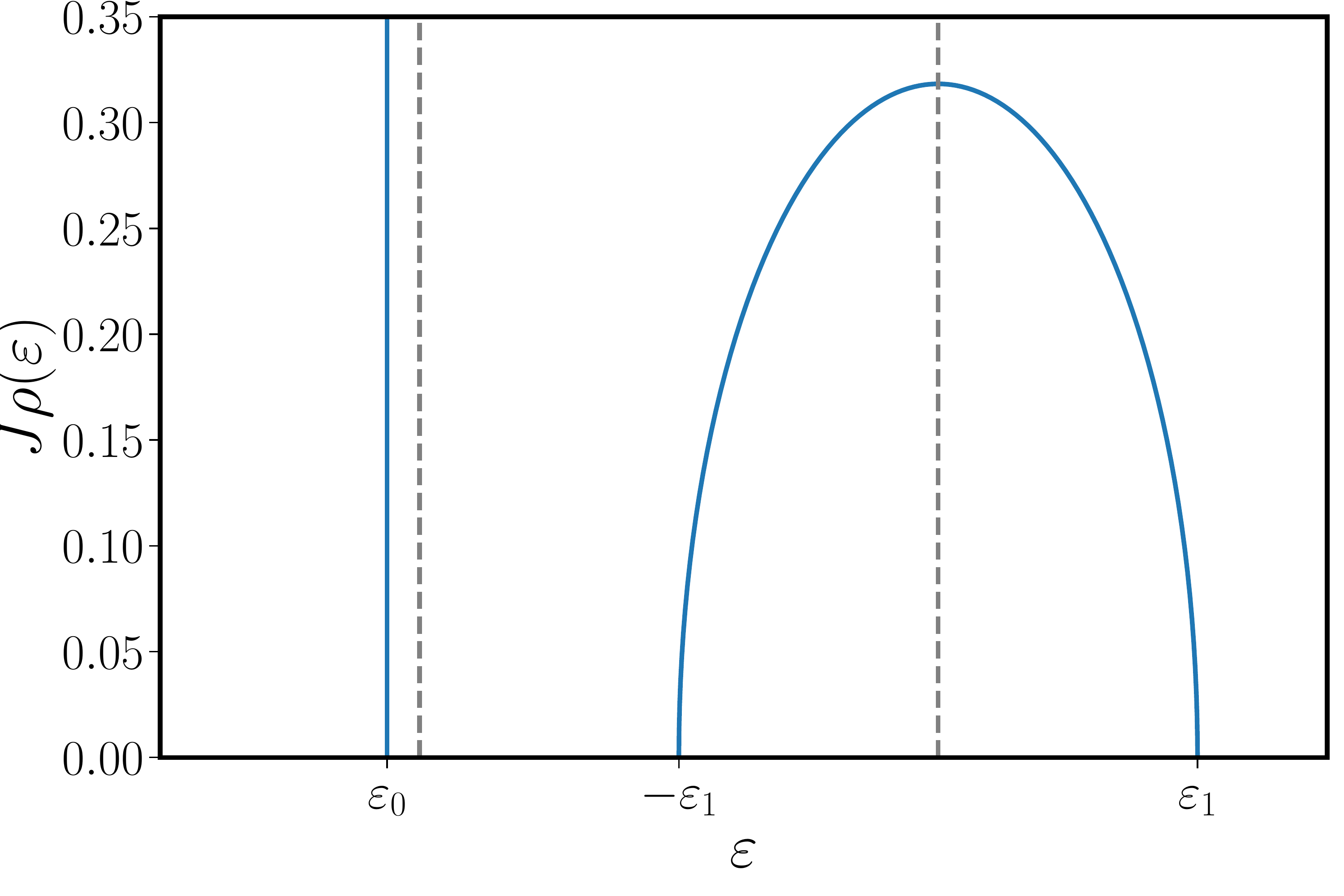}
	\caption{The density of states for the coupling matrix $U_{ij}$ in \eqref{dis_coup} is shown and compared with the clean case in \eqref{dos_clean_case} (gray dashed lines). The presence of disorder shifts the ground state energy of the small amount $J^{2}/J_{0}$ and it splits the $N-1$ degeneracy of the high energy portion of the spectrum.}
\label{Fig1app}
\end{figure}

Within this perspective, it is interesting to investigate the interplay between long-range interactions and disorder in the spectrum of quantum systems. Thus, we perturb the flat coupling matrix with small Gaussian contribution
\begin{align}
\label{dis_coup}
U_{ij}=-\frac{J_{0}}{N}-u_{ij}\quad\mathrm{with}\quad P(u_{ij})\sim \exp(-N u_{ij}^{2}/2J^{2})
\end{align}
where $J_{0}>J$.
The spectrum of the coupling matrix distributed according to \eqref{dis_coup} has been extensively studied and it is know to split in a continuous part, which obeys the celebrated semi-circular law\,\cite{metha2004random}, plus an isolated state, in analogy with the problem of a single impurity in a crystal\,\cite{kosterlitz1976spherical, edwards1976eigenvalue,dellanna2008critical}. In conclusion, the density of states for the matrix eigenvalues in the large $N$ limit reads
\begin{align}
    \rho(\varepsilon)=\rho_{0}(\varepsilon)+\frac{\delta(\varepsilon-\varepsilon_{0})}{N}\quad\mathrm{with}\quad\rho_{0}(\varepsilon)=\begin{cases} \frac{2}{\pi}\frac{\sqrt{\varepsilon_{1}^{2}-\varepsilon^{2}}}{\varepsilon_{1}^{2}}\,\,&\mathrm{if}\,\,|\varepsilon|<\varepsilon_{1}\\
    0\,\,&\mathrm{if}\,\,|\varepsilon|>\varepsilon_{1}\\
    \end{cases},
\end{align}
where $\varepsilon_{0}=-J_{0}-J^{2}/J_{0}$ and $\varepsilon_{1}=2J$. A graphical representation of the density of states is given in Fig.\,\ref{Fig1app}.

In summary, the summation in \eqref{eq:sph} can be cast into a continuous integral in the thermodynamic  limit only for $\alpha>d$, while the spectral discreteness, evidenced in the main text, causes the summation to remain discrete. The gaps in such discrete spectrum gradually reduce approaching the high-energy limit $n\to\infty$ or increasing the coupling range. In particular, in the fully connected limit $\alpha\to 0$ the entire spectrum is constituted by two levels, the non-degenerate ground state at $\varepsilon_{0}=-J_{0}$ and a $N-1$ degenerate excited state at $\varepsilon_{1}$.

In this perspective, the case of disordered fully connected couplings lies in between the short-range and the long-range cases. Indeed, at least in the weak disorder limit $J<J_{0}$, the spectrum still presents a single low-energy state with energy $\varepsilon_{0}= -J_{0}$ separated by the remaining part by a finite gap $\Delta/J_{0} = (J/J_{0}-1)^{2}$. On the other hand, the disorder lifts the degeneracy in the high energy portion of the spectrum and produces a continuous density of states in the range $\varepsilon\in [-2J,2J]$. Accordingly, the dynamical properties of the disordered model are expected to resemble the ones of a short-range interacting system rather than the ones in the strong long range regime, at least as long as the initial state is not magnetised.

\subsection{Dynamics}
\label{subsec:dyn_case}
The dynamics of the spherical model may be solved exactly as it corresponds to the dynamics of an ensemble of quantum harmonic oscillators with the addition of the constraint equation, which should be enforced at all times. However, in order to investigate how the system approaches equilibrium, we consider the simplified dynamics, where the constraint $\langle\sum_{i}\hat{s}_{i}^{2}\rangle$ is suddenly lifted at $t=0$ and the system is let free to evolve toward its unconstrained equilibrium. 

Within this perspective, the system reduces to an ensemble of free harmonic oscillators with time dependent frequency, whose dynamics may be solved analytically\,\cite{lewis1967classical,lewis1969exact,lewis1968class}. As a consequence, the dynamical state $\psi(s,t)$ in the representation of the coordinate $s$ can be expressed as
\begin{align}
\label{dyn_exp_app}
\psi(s,t)=\sum \alpha_{n}\psi_{n}(s,t),
\end{align}
where $\alpha_{n}$ are time independent constants and the dynamical eigenstates are given by
\begin{align}
\label{Dyn_Eigen_app}
\psi_{n}(x,t)=\frac{1}{\sqrt{2^{n}n!}}\left(\frac{1}{2\pi\xi^{2}(t)}\right)^{\frac{1}{4}}e^{-\Omega(t)\frac{x^{2}}{2}}
H_{n}\left(\frac{x}{\sqrt{2}\xi(t)}\right)e^{-i\left(n+\frac{1}{2}\right)\lambda(t)}.
\end{align}
The effective frequency $\Omega(t)$ as well as the overall phase $\lambda(t)$ can be expressed in terms of the effective width $\xi(t)$ 
\begin{align}
\Omega(t)=-i\frac{\dot{\xi}(t)}{\xi(t)}+\frac{1}{2\xi^{2}(t)}\quad\mathrm{and}\quad
\lambda(t)=\int^{t}\frac{dt'}{2\xi^{2}(t')}.
\end{align}
Hence, the exact time evolution of each harmonic oscillator is described by a single real function, which is the effective width $\xi(t)$ and satisfies the Ermakov-Milne equation
\begin{align}
\label{ermakov_eq_app}
\ddot{\xi}(t)+\omega(t)^{2}\xi(t)=\frac{1}{4\xi^{3}(t)}.
\end{align}

Our focus is on the dynamics induced by a sudden lift of the constraint, where the effective chemical potential of the spin waves is suddenly changed (at $t=0$) from the initial value $\mu_{i}=2\muc$ to the final value $\mu_{f}=\mu_{c}$ and the system is let free to evolve.
Within this dynamical protocol, the frequency of the eigenmodes is simply quenched from $\omi$ to $\omf$ and the solution of \eqref{ermakov_eq_app} is given by
\begin{align}
\label{xi}
\xi_{\lambda}(t) = \sqrt{1 + \epsilon_\lambda \sin[2](\omf t)} 
\end{align}
with the parameter
\begin{align}
\epsilon_k = \left( \frac{\omi}{\omf} \right)^2 - 1 \quad .
\end{align}
Due to the dissolution of the constraint the average square displacement of the spin-waves will evolve towards a expectation
\begin{align}\label{sph}
	A(t)=\frac{4}{N}\ev{\sum_i \hat{s}_i^2}= \frac{1}{N} \sum_\lambda \frac{2g}{\omega_{0}}\xi_{\lambda}(t)^{2}.
\end{align}
Inserting into the equation above the explicit solution in \eqref{xi} and performing some straightforward manipulations, one obtains
\begin{align}\label{eq:eps}
A(t) = \bar{A}-\frac{1}{N} \sum_\lambda \frac{2g}{\omi}\frac{\varepsilon_{\lambda}}{2}\cos(2\omf t)\quad\mathrm{where}\quad \bar{A}=\lim_{T\to\infty}\langle A\rangle_{T}.
\end{align}
The long time expectation of the observable $A$, i.e. $\bar{A}$, has been defined in terms of the Cesaro's average as done in Eq.\,(6) of the main text.

It is convenient to define the quantity 
\begin{align}
\chi_{A}(t)=A(t)-\bar{A}=-\frac{1}{N} \sum_\lambda \frac{g}{\omi}\varepsilon_{\lambda}\cos(2\omf t)
\end{align}
which quantifies dynamical fluctuations around the observable average. According to the definitions in closed (integrable) quantum systems, the observable $A(t)$ is said to equilibrate if the long-time Cesaro's average of its squared fluctuation vanishes 
\,\cite{reimann2008foundation,linden2009quantum,oliveira2018equilibration}
\begin{align}
\label{eq_eq}
\lim_{T\to\infty}\langle |\chi_{A}(t)|^{2}\rangle_{T}=\lim_{T\to\infty}Q_{A}(T)\approx 0
\end{align}
 Then, the equilibration of the observable $A$ (but also of all relevant physical observables in the problem) follows from the same argument outlined in \emph{Spectrum of Long-Range Systems} for the fidelity of a quantum system. Indeed, for any traslational invariant weak-long range interacting system the spectrum becomes absolutely continuous in the thermodynamic limit and the Riemann-Lebesgue lemma implies $\lim_{t\to\infty}\chi_{A}(t)\to 0$, which ensures the vanishing of dynamical fluctuations also outside the Cesaro's average.
 
More in general, for a quantum system whose initial state has no overlap on any pure point portion in the spectrum, equilibration in the sense of \eqref{eq_eq} is ensured by the Wiener's theorem\,\cite{last1996quantum}. Given these considerations, it is possible to explicitly consider the thermodynamic limit of \eqref{sph} for a strong long-range interacting system. There, increasing the system size $N$ the eigenmodes $\lambda$ of the Hamiltonian will tend to accumulate at high energy, as the spectrum becomes dense close to $\omf\sim \sqrt{2\muf}$ according to the result in Eq.\,(13) of the main text. Since most of the contribution to the summation in \eqref{sph} comes from the high-energy region with constant frequency, the dynamics of a thermodynamically large long-range system approaches the one of a single quenched harmonic oscillator, see Fig.\,4 in the main text.

In the flat interactions case ($\alpha=0$), the spectrum consists of a single infinite degenerate eigenstate, so that the dynamics of the system exactly correspond to a single harmonic oscillator, see the lower-panel in Fig.\,5. The addition of weak disorder according to \eqref{dis_coup} completely disrupts the above picture as it restores spectral continuity\,\cite{kosterlitz1976spherical, edwards1976eigenvalue,dellanna2008critical}, yielding
\begin{align}
\label{dis_res}
\chi_{A}(t)=-\int_{-2J}^{2J}\frac{2g\varepsilon_{s}\rho(s)}{\sqrt{2\mu_{i}-s}}\cos(2\omega_{f,s}t)ds.
\end{align}
Since $\mu_{i},\muf> 2J$ for any state in the system, the integrand in \eqref{dis_res} is continuous. Hence, the Riemann-Lebesgue lemma is applicable, leading to the conclusion that $\lim_{t\to\infty}\chi_{A}(t)\to 0\quad \forall J>0$. The consequence of this phenomenon are clearly visible in the upper panel of Fig.\,5 in the main text, where the observable $A(t)$ is found to equilibrate for any finite disordered strength.

\begin{figure*}[ht]
\centering
\includegraphics[width=.45\linewidth]{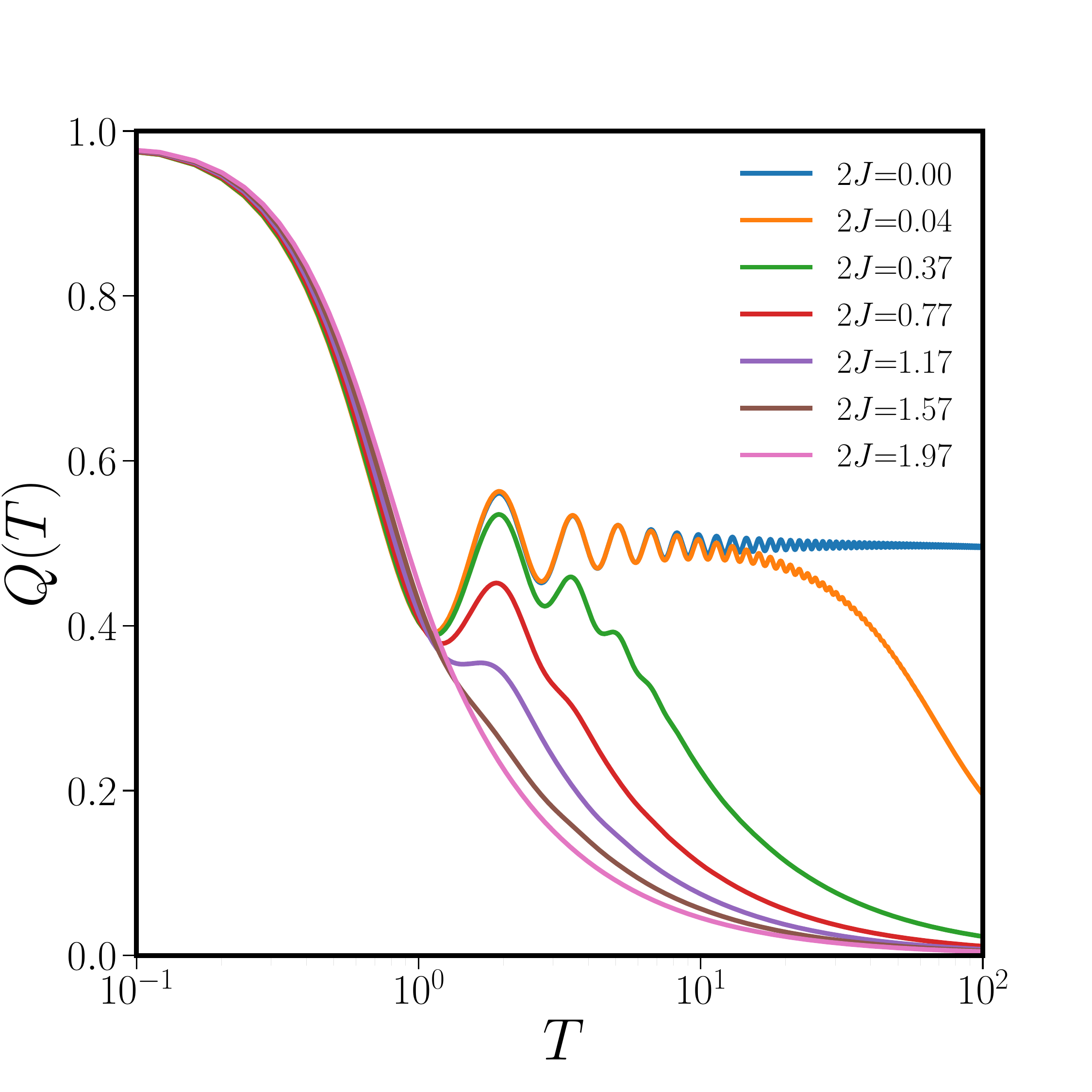}\quad
\includegraphics[width=.45\linewidth]{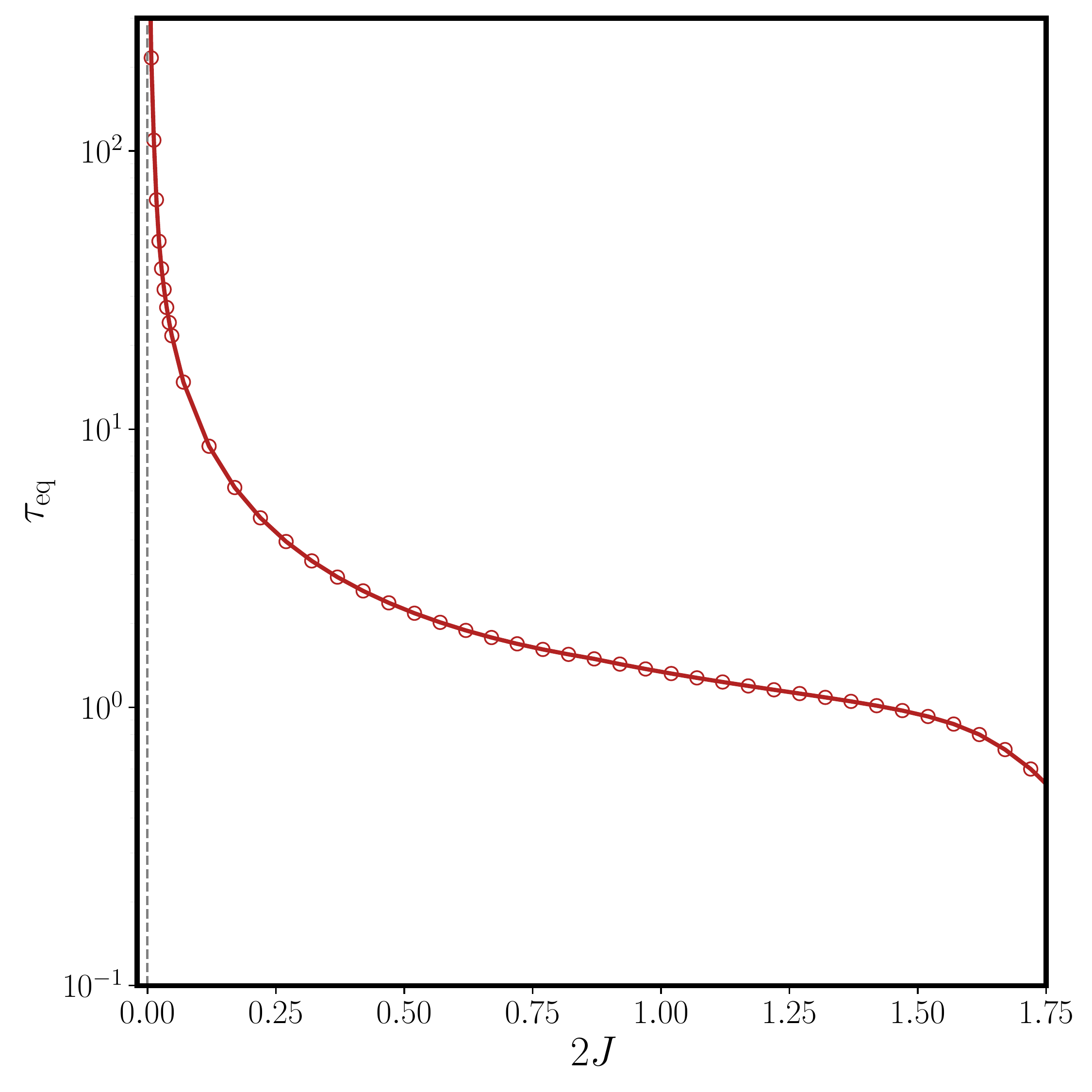}
\caption{Left panel: the average width of dynamical fluctuations for a fully connected system with Gaussian random couplings, after a quench from the equilibrium state at $\mu_{i}=2\muc$ to $\muf=\muc$. The equilibration of the system is clearly visible from the exponential decay of this quantity. As the intensity of disorder is reduced, the exponential tails shift to longer times, until it finally disappears in the $2J\to0$ limit, where no relaxation is possible, due to the discrete nature of the spectrum. Right panel: in agreement with the previous discussion, the equilibration time of the system defined by \eqref{eq_time} diverges as the system approaches the zero disorder limit. \label{Fig2app}}
\end{figure*}

Interestingly, the decay of the amplitude of dynamical fluctuations for disordered systems is exponential, leading to the definition of equilibration time $\tau_{\rm eq}$
\begin{align}
\label{eq_time}
Q_{A}(T)\approx R e^{-T/\tau_{\rm eq}}.
\end{align}
A numerical analysis as a function of the disorder strength $J$ reveals that, as expected, the equilibration time $\tau_{\rm eq}$ is monotonically decreasing as a function of the disorder strength, see Fig.\,\ref{Fig2app}.

Despite the profound differences between the definition of equilibration in the classical and quantum realms, the results depicted in this section are remarkably similar to the one obtained for a classical spherical model with disordered couplings distributed according to \eqref{dis_coup}. Indeed, the Langevin dynamics of this model do not display any signatures of metastability at least as long as the initial state is not magentized, see Chap.\,4 of Ref.\,\cite{cirano2006random}.

The present analysis has been pursued with the scope of characterising the long-time equilibration dynamics of many-body quantum systems. Therefore, the thermodynamic limit has generally been taken before the long-time limit. However, the same conclusion could have been obtained considering the long time limit average of dynamical fluctuations for a finite system, which yields\,\cite{reimann2008foundation,linden2009quantum,oliveira2018equilibration}
\begin{align}
\label{fs_eq}
\lim_{T\to\infty}Q_{A}(T)\propto \frac{1}{d_{\rm eff}}
\end{align}
where $d_{\rm eff}$ roughly corresponds to the number of modes participating to the dynamics (it should be bare in mind that for a finite system the entire spectrum is discrete and only a finite number of modes exist). As the thermodynamic limit is approached, the spectrum becomes continuous and eigenvalues becomes dense in any arbitrarily small energy range. Then, in the continuous limit one shall have $d_{\rm eff}\to\infty$ for any dynamics involving initial states in the continuous spectrum. On the other hand, for long-range systems with $\alpha<d$ no continuous theory may be defined, as the only dense region in the spectrum occurs around the energy maximum, where infinitely many degenerate eigenvalues are forming,  violating the assumption of non-degenerate energy gaps at the foundations of \eqref{fs_eq}\,\cite{short2011equilibration}.
\bibliography{bibl.bib}

\end{document}